\documentclass{aa}  
\usepackage{graphicx}
\usepackage{cancel}
\usepackage[colorlinks,allcolors=blue]{hyperref}
\bibpunct{(}{)}{;}{a}{}{,}
\usepackage{natbib}
\usepackage[varg]{txfonts}
\usepackage{tabularx} 
\usepackage{graphicx}
\usepackage{txfonts}

\def\apjl{\rm ApJL}
\def\apjs{\rm ApJS}

\def\aap{\rm A\&A}

\usepackage{url}
\begin{document} 

   \title{2D magnetohydrodynamic jet simulations: Properties of recollimation shocks}

 \author{S. Boula
          \inst{1}\orcid{0000-0001-7905-6928}, F. Tavecchio\inst{1}\orcid{0000-0003-0256-0995}, G. Bodo\inst{2}\orcid{0000-0002-8913-5176} , N.Vlahakis\inst{3}\orcid{0000-0002-9265-4081}\and  P. Coppi\inst{4}\orcid{0000-0001-9604-2325}
        }

   \institute{INAF – Osservatorio Astronomico di Brera, Via E. Bianchi 46, I-23807 Merate, Italy\\
              \email{styliani.boula@inaf.it}
\and 
            INAF, Osservatorio Astrofisico di Torino, Strada Osservatorio 20, I-10025 Pino Torinese, Italy
               \and
         Department of Physics, National and Kapodistrian University of Athens, GR 15783 Zografos, Greece
\and 
Department of Astronomy, Yale University, PO Box 208101, New Haven, CT 06520-8101, USA
 }
 \date{Received 16/06/2026; accepted 06/08/2026}
  \abstract
{Recollimation shocks are a frequent outcome in overpressured relativistic jets and are crucial for interpreting stationary features in active galactic nuclei (AGNs). The precise influence of magnetic fields on jet stability, energy dissipation, and variability remains debated, particularly as different field configurations can significantly alter shock properties and the onset of fluid instabilities.}
 {We perform a study of 2D axisymmetric relativistic magnetohydrodynamic (RMHD) jets to quantify how the ambient density contrast ($\nu$), pressure ratio ($P$), magnetization ($\sigma$), and magnetic pitch parameter ($\alpha$) govern the formation and strength of the first recollimation shock. We also assess how these parameters create the local geometric conditions favorable for the centrifugal instability (CFI), utilizing linear theory as a diagnostic.}
 {We used high-resolution 2D RMHD simulations of conical jets propagating into a uniform external medium. We analyzed the resulting steady-state profiles to evaluate magnetic field partitioning and flow dynamics, and compare them against linear stability criteria to diagnose the likelihood of CFI development.}
 {We find that the jet's global geometry is affected by the magnetic pressure. The recollimation distance decreases monotonically with increasing magnetization, $\sigma$, as increased magnetic forces immediately limit jet expansion. Remarkably, in the magnetically dominated regime, the ratio of the magnetized recollimation distance ($z_{\rm MHD}$) to its purely hydrodynamic counterpart ($z_{\rm HD}$) converges on a power-law scaling, $z_{MHD}/z_{HD} \propto (B_0^2/P_{ext})^{-1/3}$, where $B_0$ is the initial magnetic field and $P_{ext}$ the external pressure. This demonstrates that the physics of recompression is robust to variations in the jet's density contrast and pressure ratio. Jets with a high density contrast relative to the ambient medium or a high internal pressure further enhance field compression. Furthermore, synthetic synchrotron maps show that a dominant toroidal field yields highly boosted, localized emission knots, whereas a strong poloidal field creates a diffuse profile and shifts the recollimation zone downstream. Regions susceptible to CFI are determined primarily by the local $\sigma_{\text{tor}}/\Gamma^2$ profile and streamline curvature created during recollimation.}
 {}
 \titlerunning{Recollimation shocks in 2D MHD jets}
 
   \authorrunning{Boula et al.}
   \maketitle

\section{Introduction}
Relativistic jets from active galactic nuclei (AGNs) are highly energetic outflows that extend from subparsec scales near the supermassive black hole to megaparsec distances, producing nonthermal emission across the electromagnetic spectrum and potentially contributing to high-energy neutrinos and ultrahigh-energy cosmic rays \citep[e.g.,][]{Blandford19,SRT09,Petropoulou2020,BM2022}. Despite preserving their collimation over large scales \citep{Willis1974}, jets are subject to a range of hydrodynamic and magnetohydrodynamic instabilities \citep[for a review, see][and references therein]{KP2021}, including Kelvin–Helmholtz, Rayleigh–Taylor, centrifugal, current-driven, and pressure-driven modes \citep[e.g.,][]{Bodo1989, Bodo2004,Bodo2013, Bodo2019,GK18b,Das2019,wang2023,Bodo2021, Musso2024, Boula25}. Observations of nearby radio galaxies, such as M87 and BL Lac, show jets that expand and accelerate near the core before gradually collimating at larger distances \citep{Cohen2014,Walker2018}. This evolution is often associated with recollimation shocks arising from pressure imbalances between the jet and its environment \citep{KF1997,NS2009,BT2018,GK2018,CBT2024,Boula25, costa26,Elley2026,Richards26}. 

Recollimation shocks influence jet morphology and can serve as localized sites of particle acceleration \citep[e.g., a recent study of shock particle acceleration,][]{Shirin25}. They may also be linked to high-energy variability, as in Mrk 421, where multiple stationary shocks have been proposed to explain complex X-ray flaring \citep{Hervet2019} or show quasi-periodic emission \citep{Hu25}. These observations underscore the importance of parsec-scale recollimation zones in determining the dynamical and radiative properties of jets. Furthermore, while theoretical models rely heavily on the jet's fundamental magnetization ($\sigma$), constraining $\sigma$ on parsec scales remains a major observational challenge. Current estimates largely depend on complex polarization measurements or Very-Long-Baseline Interferometry (VLBI) core-shift analyses \citep[e.g.,][]{Pushkarev2012, Kravchenko2020}. Identifying observable geometric proxies for magnetization, such as the spatial scaling of stationary recollimation knots, is therefore important for linking relativistic magnetohydrodynamic (RMHD) theory directly to high-resolution jet observations.

Early axisymmetric models established the purely hydrodynamic baseline, predicting a stable sequence of recollimation and reflection shocks \citep{KF1997}, while subsequent studies extended these structures to magnetized flows \citep[e.g.,][]{Mizuno2015,PK2015,2016Fromm}. However, 3D studies show that such structures can become unstable beyond the first shock. The centrifugal instability (CFI), driven by streamline curvature in the recollimation region, is a key mechanism. The CFI is inherently 3D and can develop even in the absence of a density contrast between the jet and the ambient medium \citep{GK2018, Boula25}. An accurate characterization of the quasi-steady axisymmetric base flow is therefore essential, as it sets the local geometric and magnetic conditions governing instability growth.

Magnetic fields introduce additional complexity \citep{GK2018,Komissarov2019,Gottlieb2020,Matsumoto2021,Hu2024,Boula25}. Fundamentally, the toroidal magnetic field plays a well-established role in the asymptotic collimation and initial acceleration of relativistic outflows \citep[e.g.,][]{Komissarov2007, Lyubarsky2009, Tchekhovskoy2011}. However, while the global magnetic confinement mechanism is widely recognized, its precise influence on the local geometry of the first recollimation shock, and how this magnetic pressure competes with internal thermal expansion across different jet regimes, remains less thoroughly mapped. In these magnetized recollimation zones, current- and pressure-driven instabilities interact with hydrodynamic modes, and the field’s configuration and strength are critical for stability. Linear analyses have provided quantitative estimates of growth rates under different magnetic setups \citep{Bodo2013, Bodo2019,Kim18,Vlahakis2023,Vlahakis2024, Musso2024}. The interplay among magnetization, velocity shear, density, and external pressure gradients ultimately determines the geometry of the first recollimation shock and the conditions for the onset of instability.

In this work, we perform a systematic study to quantify how the ambient density contrast ($\nu$), pressure ratio ($P$), magnetization ($\sigma$), and magnetic pitch parameter ($\alpha$) govern the topology, spacing, and strength of recollimation shocks, particularly across the transition from hydrodynamic to magnetically dominated regimes. Furthermore, we examine the role that local streamline curvature induced by these shocks plays in triggering the CFI. 
Finally, we assess whether the onset of such instabilities can be robustly predicted by applying linear stability analysis to the evolved, quasi-steady jet profiles. We perform high-resolution 2D axisymmetric RMHD simulations and combine them with linear analysis. This approach allows us to quantify how parameters such as magnetization, density contrast, and magnetic pitch shape the structure of recollimation shocks and to establish the geometric conditions for CFI growth, while the linear analysis provides the corresponding instability growth rates.

The paper is organized as follows. Section~\ref{sec:methods} presents the numerical setup and the linear stability diagnostic. Section~\ref{sec:fiducial} establishes the baseline physics using a fiducial model. Section~\ref{sec:parameter} provides a systematic parameter study of the recollimation shock geometry. Section~\ref{sec:dynamics} details the global jet dynamics and the effects of magnetic pitch on the flow morphology. Section~\ref{sec:cfi} analyzes the stability of the evolved flows against the CFI. Finally, we discuss our findings in Sect.~\ref{sec:discussion} and conclude in Sect.~\ref{sec:conclusions}.

\section{Numerical setup and methods}
\label{sec:methods}

Observations indicate that quasi-stationary features, likely linked to recollimation shocks, are present on parsec scales within AGN jets \citep[see for example][]{Paraschos2025,Hu25}. To identify the steady-state solutions characterized by a distinct recollimation structure, we performed 2D axisymmetric simulations, following the method detailed in \cite{Boula25}.

\subsection{RMHD equations and grid}
The numerical simulations were carried out by solving the equations of relativistic ideal magnetohydrodynamics in 2D, using the ideal RMHD module of the PLUTO code \citep{PLUTO2007}.  The equations were evolved with second-order Runge-Kutta time stepping, linear reconstruction, and the Harten-Lax-van Leer-Discontinuities (HLLD) Riemann solver \citep{Mignone2009}. Beyond these methods, magnetic field evolution required additional care. Constrained transport \citep{Balsara1999, Londrillo2004} was used to maintain the condition $\nabla \cdot \mathbf{B} = 0$.

The system of equations in the conservation form reads
\begin{equation}
\frac{\partial}{\partial t}
\begin{pmatrix}
D \\
\mathbf{m} \\
E_t \\
\mathbf{B}
\end{pmatrix}
+ \nabla \cdot 
\begin{pmatrix}
D\mathbf{u} \\
w_t \Gamma^2 \mathbf{uu} - \mathbf{bb} + \mathbf{I} p_t \\
\mathbf{m} \\
\mathbf{u B} - \mathbf{B u}
\end{pmatrix}
=
\begin{pmatrix}
0 \\
\mathbf{f_g} \\
\mathbf{u \cdot f_g }\\
0
\end{pmatrix}\label{eq:1}
.\end{equation}
We adopted units in which the speed of light is $c=1$, and the magnetic field, ${B}$, is rescaled to absorb the factor $\sqrt{4\pi}$. The conserved variables and the Taub-Matthews equation of state \citep{Mignone2005} are defined in Appendix \ref{app:3dres}.

The setup of each 2D simulation uses cylindrical coordinates $(r, z)$, with a domain of 
$[0, L_{r,2D}] \times [1, L_{z,2D}]$. The 2D simulations use a grid of $1400 \times 2200$ zones. In the central region  $[0, 2] \times [1, 20]$, the grid is uniform, providing a resolution of 50 points per initial jet radius, while the outer regions are geometrically stretched. 

\subsection{Initial conditions and parameters}
A relativistic conical jet with an initial opening angle, $\theta_j$, is injected into the computational domain at a distance, $z_0$, from the cone's apex, which is embedded in the confining ambient medium. 
Our simulations adopt a jet with an initial opening angle of $\theta_j = 0.1$ and a Lorentz factor of $\Gamma = 10$. The total jet power ($L_{\text{jet}}$) is defined as the sum of its kinetic (hydrodynamic) and magnetic (Poynting) components:
\begin{equation}
L_{\text{jet}} = \pi z_0^2 \theta_j^2 \Gamma_j^2 v_j\left( \mathcal{C}_{HD}\rho_j h_j + \mathcal{C}_{MHD}{B_0^2} \right)
,\end{equation}
where $\mathcal{C}_{HD} \simeq 0.5$ and $\mathcal{C}_{MHD}\simeq 0.5$ are the factors from the integrated radial profiles. We assume $z_0=0.03 ~\rm{pc}$ and $\rho_{ext,0}=10^3 ~\rm{m_p cm}^{-3}$. The initial normalization for the magnetic field is $B_0$,  and represents a helical structure with $\alpha=1 $ (Appendix \ref{app:mag_init}).
The key parameters governing the system are the density ratio, $\nu = {\rho_j}/\rho_{ext}$, the pressure ratio, $P_{\text{ratio}} = P_{j}/P_{ext}$, and the cold magnetization, $\sigma ={B_0^2}/{\rho(z_0)}$; see Tab. \ref{tab:parameters} for details.
We assume a decreasing power-law profile for the ambient pressure with an index of $\eta = 0.5$. The transverse transition of all variables was smoothed using hyperbolic secant profiles to avoid numerical noise (see Appendix \ref{app:smooth}). At the base of the jet we imposed a magnetothermal equilibrium condition. Furthermore, we defined the hot magnetization, $\sigma_{\rm hot}= {B_0^2}/{\rho(z_0) h}$, the poloidal component, $\sigma_{\rm pol}={B_{\rm pol}^2}/{\rho h}$, and the toroidal component, $\sigma_{\rm tor}={B_{\phi}^2}/{\Gamma^2\rho h}$, where $B_{\rm pol}$ and $B_{\phi}$ are the poloidal and toroidal components of the magnetic field, respectively. We defined the magnetic pressure, $P_B = {B^2}/{\Gamma^2}+{(\mathbf{u}\cdot \mathbf{B})^2}/{2} $, and the temperature, $T=P/\rho$.

\subsection{Linear stability method}
To assess the stability of the simulated jets against CFI, we employed a linear stability analysis as a diagnostic tool. We mapped the steady-state 2D profiles to the linear stability criterion derived for a magnetized, rotating cylindrical shell, as detailed in \cite{Vlahakis2023} and \cite{Boula25}. The growth of the CFI is linked to the ratio $\sigma_{\mathrm{tor}}/\Gamma^2$, which measures the relative strength of magnetic tension to inertia. We utilized this ratio to identify regions in the evolved flow that are susceptible to instability.

\section{Anatomy of a magnetized recollimation shock }
\label{sec:fiducial}
To establish a clear physical baseline for understanding the complex interplay between jet hydrodynamics and magnetic field structure, we first analyzed a single representative fiducial case in detail. We selected the moderately magnetized, pressure-matched light jet (Case C3: $\sigma=1$, $\nu=10^{-5}$, $P_{\rm ratio} \sim 1$) as our standard reference model. Figure~\ref{fig:fiducial_combined} provides a comprehensive overview of the steady-state structure of this jet, combining the axial profiles of pressure and Lorentz factor, 2D morphology, and the 1D radial pressure profiles driving its confinement. As shown by the map in the rightmost central panel, the plasma parameter, $\beta=P_{\rm th}/P_{\rm B}$, associated with the jet is close to 1, indicating that the dynamics is ruled by the combination of hydrodynamical and magnetic forces.

\subsection{Morphology and energy conversion}
The macroscopic structure of the jet is illustrated in the central panels of Fig.~\ref{fig:fiducial_combined}, which map the decimal logarithm of rest-mass density ($ \rho$), thermal pressure ($ P_{th}$), the bulk Lorentz factor ($\Gamma$), and plasma, $\beta$. A characteristic bounding streamline (dashed white line) tracks the expansion and subsequent contraction of the jet core. 

Upon injection, the jet undergoes expansion, the local thermal energy is converted into bulk kinetic energy. This conversion is explicitly visible in the leftmost panel (axial distance versus normalized profiles): tracking the flow along the central axis ($r=0$), the normalized thermal pressure ($P_{th}/P_{th_0}$) drops by an order of magnitude, while the normalized Lorentz factor ($\Gamma/\Gamma_0$) climbs steadily, reaching a peak velocity just prior to $z \approx 3.0$. In the 2D maps, this acceleration region corresponds to the bright, expanding bright core in the $\Gamma$ panel and the dark, low-pressure cavity in the $P_{th}$ panel.

The expansion is ultimately halted by the confining environment. A curved recollimation shock is created at which the streamlines are bend toward the axis. The recollimation shock reaches the axis at$z \approx 3.0$, where the flow experiences an abrupt discontinuity: the Lorentz factor plummets, and the density and thermal pressure spike drastically as kinetic energy is violently dissipated back into internal energy (far left panel of Fig. \ref{fig:fiducial_combined}). The point of maximum radial expansion after the recollimation shock ($r_{max} \approx 0.11$) occurs at an axial distance of $z_{max} = 2.20$ (central panels of 2D maps of Fig. \ref{fig:fiducial_combined}), marked by the cyan cross. After passing $z_{max}$, the jet boundary curves inward, driving a convergent flow that culminates in a strong reflection shock on the axis at $z \approx 3.0$.

\subsection{Radial pressure distribution and reconfinement}
\begin{figure*}[htbp]
\begin{center}
   \includegraphics[width=0.95\linewidth]{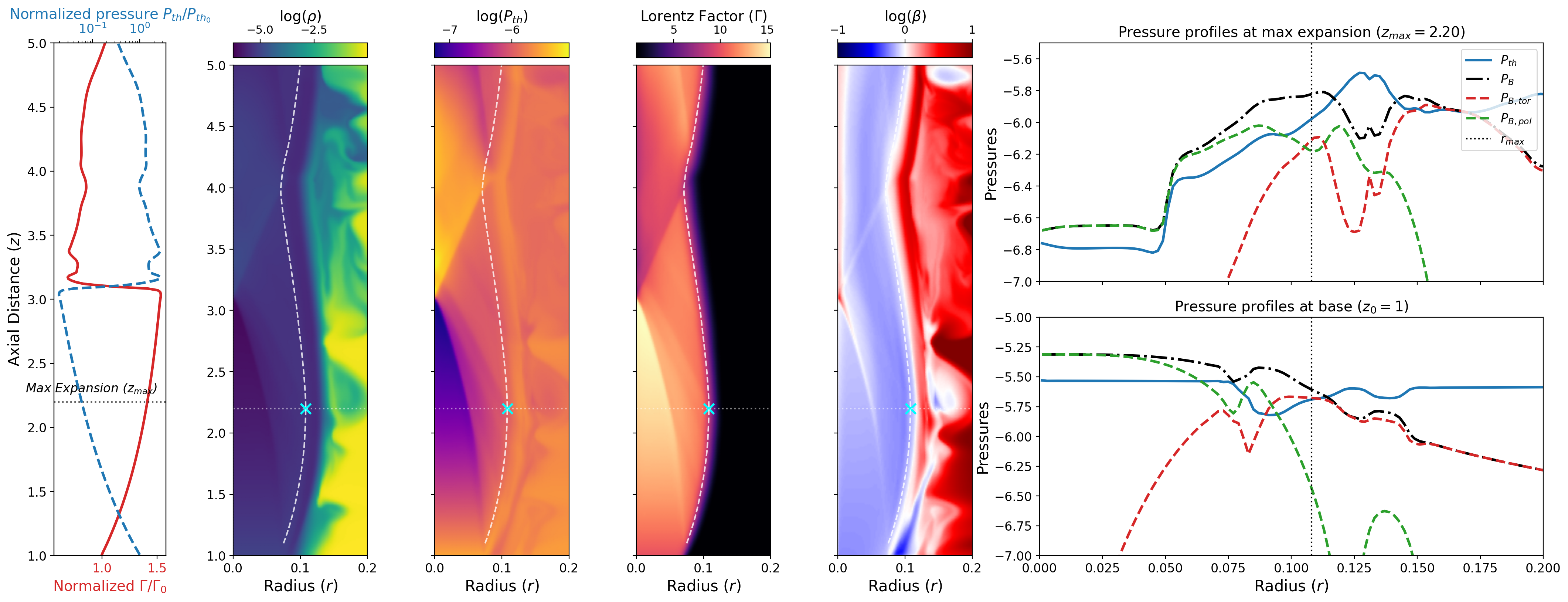} 
   \caption{Anatomy of the fiducial magnetized jet (Case C3: $\sigma=1$, light jet). {\textit{Far left}:} 1D axial profiles along $r=0$, showing the conversion of internal energy ($P_{th}$; dashed blue) to bulk kinetic energy ($\Gamma$; solid red), followed by a sharp recollimation shock at $z \approx 3.0$. {\textit{Center columns}:} 2D maps of density, thermal pressure, Lorentz factor, and plasma, $\beta$. The dashed white line is a representative streamline marking the jet boundary, with the point of maximum radial expansion ($z_{max}=2.20$, $r_{max} \approx 0.11$) denoted by a cyan cross. {\textit{Far fight}:} 1D radial pressure profile, the thermal pressure ($P_{th}$; solid blue), the total magnetic pressure ($P_B$; dash-dotted black), the toroidal component of the magnetic pressure ($P_{B,tor}$; dashed red), and the poloidal component of the magnetic pressure ($P_{B,pol}$; dashed green). The bottom panel depicts the pressure profiles at the jet's base.}
   \label{fig:fiducial_combined}
\end{center}
\end{figure*}

To understand the physical mechanism driving recollimation, it is highly instructive to examine the local pressure distribution across the jet slice. The rightmost column of Fig.~\ref{fig:fiducial_combined} details the radial pressures extracted precisely at the slice of maximum radial expansion ($z=2.20$) and at the base of the jet ($z=1$).

The confinement is dictated by the profiles of the thermal pressure ($P_{th}$; solid blue line) and the magnetic pressure  ($P_B$; dashed black line). As noted above, $\beta$ is close to one for the parameters considered here, suggesting that both thermal expansion and magnetic confinement are equally vital.  
The net radial force dictating streamline curvature can be approximated by the radial balance equation:
\begin{equation}
F_r \approx - \frac{\partial}{\partial r} \left( P_{\rm th} + \frac{B^2}{2} \right) - \frac{B_\phi^2}{\Gamma^2r},
\label{eq:radial_force}
\end{equation}
where the term $B_\phi^2/\Gamma^2 r$ represents the azimuthal hoop stress. Contrary to intuitive expectations that the magnetic confinement is mainly driven by this magnetic tension, our analysis shows that the inward total pressure gradient (comprising both thermal and magnetic components) plays an equally vital role. Under the combined effect of this boundary pressure gradient and the toroidal field's tension.

At the jet base (bottom panel), the pressure profiles strongly favor outward expansion. The thermal pressure remains relatively uniform across the inner jet, while the total magnetic pressure -- initially dominated by its poloidal component near the axis -- decreases toward the periphery ($r \approx 0.10$). This lack of a strong bounding pressure gradient at the injection scale drives the initial lateral expansion of the flow. 
At higher altitudes (upper panel), the upstream region shows an overall decrease compared to the base case due to adiabatic expansion and the $1/r^2$ decrease in the magnetic field. Downstream of the shock, both the internal and magnetic pressures increase from the axis toward the external medium, producing an effective confining force that constrains the lateral expansion. This analysis establishes our physical baseline: in moderately magnetized jets ($\sigma=1$, $\beta \simeq 1$), the recollimation process is driven by downstream buildup of the magnetic and thermal pressure gradients, while the confining tension of the toroidal field lines plays a secondary role. In the following sections, we explore the effect of the modification of this baseline magnetization and magnetic structure.

\section{Parameter study}
\label{sec:parameter}
We now perform a systematic study to quantify how the ambient density contrast, pressure ratio, and magnetization govern the shock properties. Table \ref{tab:parameters} lists the simulation families. The density maps of these cases are presented in Appendix \ref{app:density_map}.

\subsection{Steady-state configurations}
The 2D steady configurations obtained from our simulations display the characteristic structure of relativistic jets confined by an external medium. Each jet evolves from the injection conditions toward a quasi-steady equilibrium, in which the balance among the jet's internal pressure, magnetic pressure gradients, and external confinement establishes a sequence of expansion and recollimation regions.

\begin{table}[h!]
    \centering
    \caption{Simulation families, with each case corresponding to a specific combination of density ratio ($\nu$), pressure ratio ($P_{ratio}$), and magnetization ($\sigma$). $L_{jet}$ is the jet's power.}
\begin{tabular}{c cccccc}
\hline\hline
\textbf{Case} & $\nu$ & $P_{\rm ratio}$ & $\sigma$ & $\Gamma$ & $\theta_j$ & $\log L_{\rm jet}$ (erg/s) \\ \hline
A1 & $10^{-5}$ (light) & $10^{-3}$ & 0   & 10 & 0.1 & 42.68 \\ 
A2 & $10^{-5}$ (light) & $10^{-3}$ & 0.1 & 10 & 0.1 & 42.95 \\ 
A3 & $10^{-5}$ (light) & $10^{-3}$ & 1   & 10 & 0.1 & 43.26 \\ 
A4 & $10^{-5}$ (light) & $10^{-3}$ & 3   & 10 & 0.1 & 43.45 \\ \hline

B1 & $10^{-4}$ (heavy) & $10^{-3}$ & 0   & 10 & 0.1 & 43.68 \\ 
B2 & $10^{-4}$ (heavy) & $10^{-3}$ & 0.1 & 10 & 0.1 & 43.78 \\ 
B3 & $10^{-4}$ (heavy) & $10^{-3}$ & 1   & 10 & 0.1 & 43.95 \\ 
B4 & $10^{-4}$ (heavy) & $10^{-3}$ & 3   & 10 & 0.1 & 44.08 \\ \hline

C1 & $10^{-5}$ (light) & $\sim 1$ & 0   & 10 & 0.1 & 42.99 \\ 
C2 & $10^{-5}$ (light) & $\sim 1$ & 0.1 & 10 & 0.1 & 43.14 \\ 
C3 & $10^{-5}$ (light) & $\sim 1$ & 1   & 10 & 0.1 & 43.36 \\ 
C4 & $10^{-5}$ (light) & $\sim 1$ & 3   & 10 & 0.1 & 43.52 \\ \hline

D1 & $10^{-4}$ (heavy) & $\sim 1$ & 0   & 10 & 0.1 & 43.72 \\ 
D2 & $10^{-4}$ (heavy) & $\sim 1$ & 0.1 & 10 & 0.1 & 43.82 \\ 
D3 & $10^{-4}$ (heavy) & $\sim 1$ & 1   & 10 & 0.1 & 43.98 \\ 
D4 & $10^{-4}$ (heavy) & $\sim 1$ & 3   & 10 & 0.1 & 44.10 \\ \hline

E1 & $10^{-5}$ (light) & $10$ & 0   & 10 & 0.1 & 43.69 \\ 
E2 & $10^{-5}$ (light) & $10$ & 0.1 & 10 & 0.1 & 43.72 \\ 
E3 & $10^{-5}$ (light) & $10$ & 1   & 10 & 0.1 & 43.79 \\ 
E4 & $10^{-5}$ (light) & $10$ & 3   & 10 & 0.1 & 43.86 \\ \hline

F1 & $10^{-4}$ (heavy) & $10$ & 0   & 10 & 0.1 & 43.99 \\ 
F2 & $10^{-4}$ (heavy) & $10$ & 0.1 & 10 & 0.1 & 44.04 \\ 
F3 & $10^{-4}$ (heavy) & $10$ & 1   & 10 & 0.1 & 44.14 \\ 
F4 & $10^{-4}$ (heavy) & $10$ & 3   & 10 & 0.1 & 44.23 \\ \hline
\end{tabular}
    \label{tab:parameters}
\end{table}

\begin{figure}[htbp]
\begin{center}
   \includegraphics[width=\linewidth]{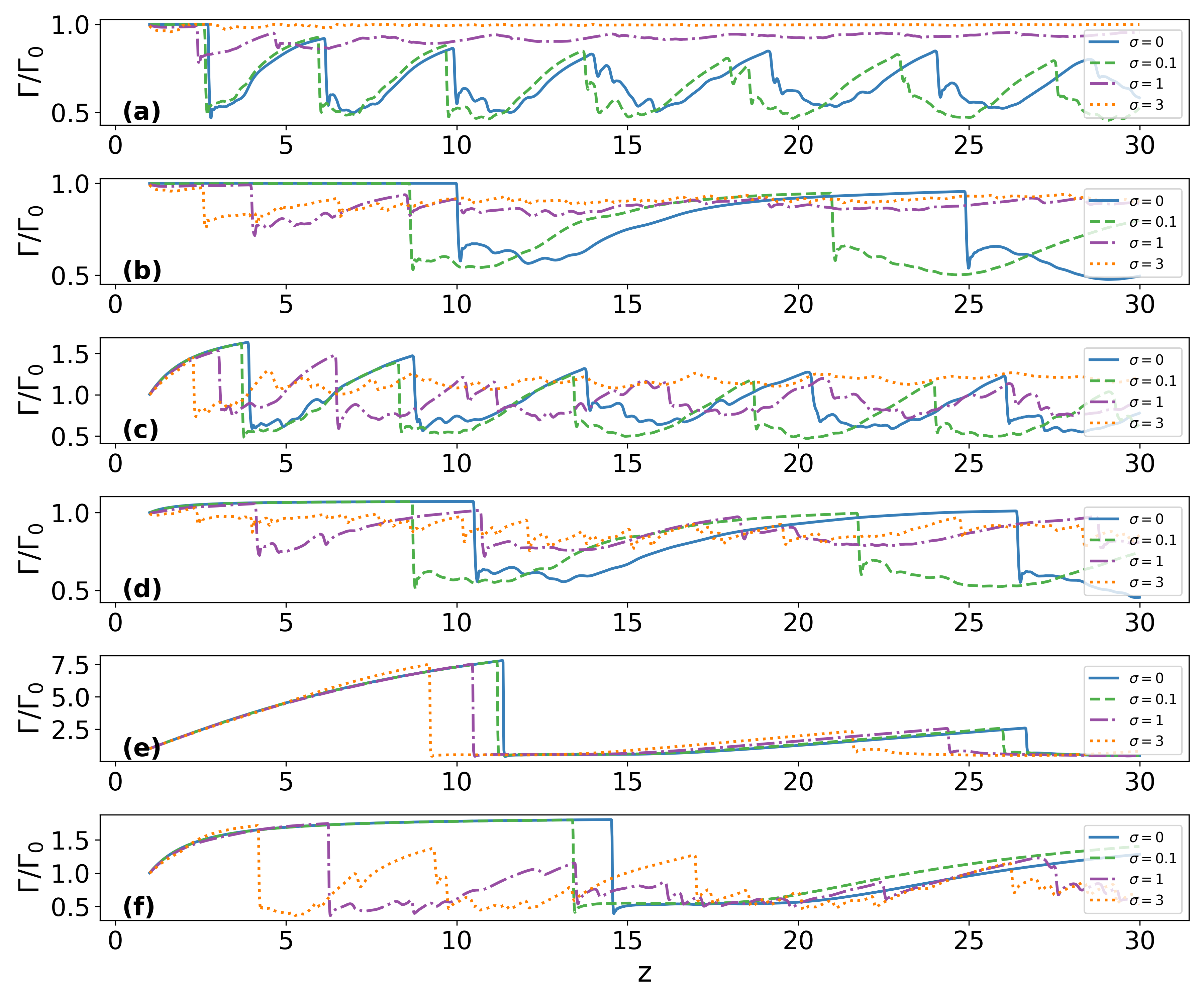}
\caption{Axial profiles of the Lorentz factor normalized to its injection value, $\Gamma / \Gamma_{0}$, as a function of the distance $z$ along the jet axis. The different panels correspond to different model families (see Table \ref{tab:parameters}). Each line refers to a different magnetization at injection: $\sigma=0$ (solid blue), $\sigma=0.1$ (dashed green), $\sigma=1$ (dotted orange), and $\sigma=3$ (dash-dotted purple). \label{fig:Gamma_profiles}}
\end{center}
\end{figure}

Figure~\ref{fig:Gamma_profiles} shows the axial profiles of the normalized Lorentz factor for the simulated families across varying magnetizations (also in Appendix \ref{ap:parameters} comparisons across other parameters are presented). 
In the hydrodynamic limit ($\sigma=0$), the behavior follows the standard model described in \cite{KF1997}, forming  a recollimation shock.

The details of the dynamic depend on the jet's thermal content. For “cold” jets (Runs A and B),  the normalized temperature is small ($T \ll 1$) and there is no conversion of thermal energy into bulk kinetic energy of the jet. Consequently, the bulk Lorentz factor is constant until the flow reaches the shock and it is suddenly decelerated.  

Conversely, the “hot” jets (Run E) exhibit a fundamentally different behavior. These models are characterized by very high internal pressure ($T \gg 1$; specifically $\sim 10$ for Run E). In this thermally dominated regime,  the flow can convert the internal pressure into kinetic energy, leading to the acceleration of the flow before the shock.

\begin{figure}[htbp]
\begin{center}
\includegraphics[width=\linewidth]{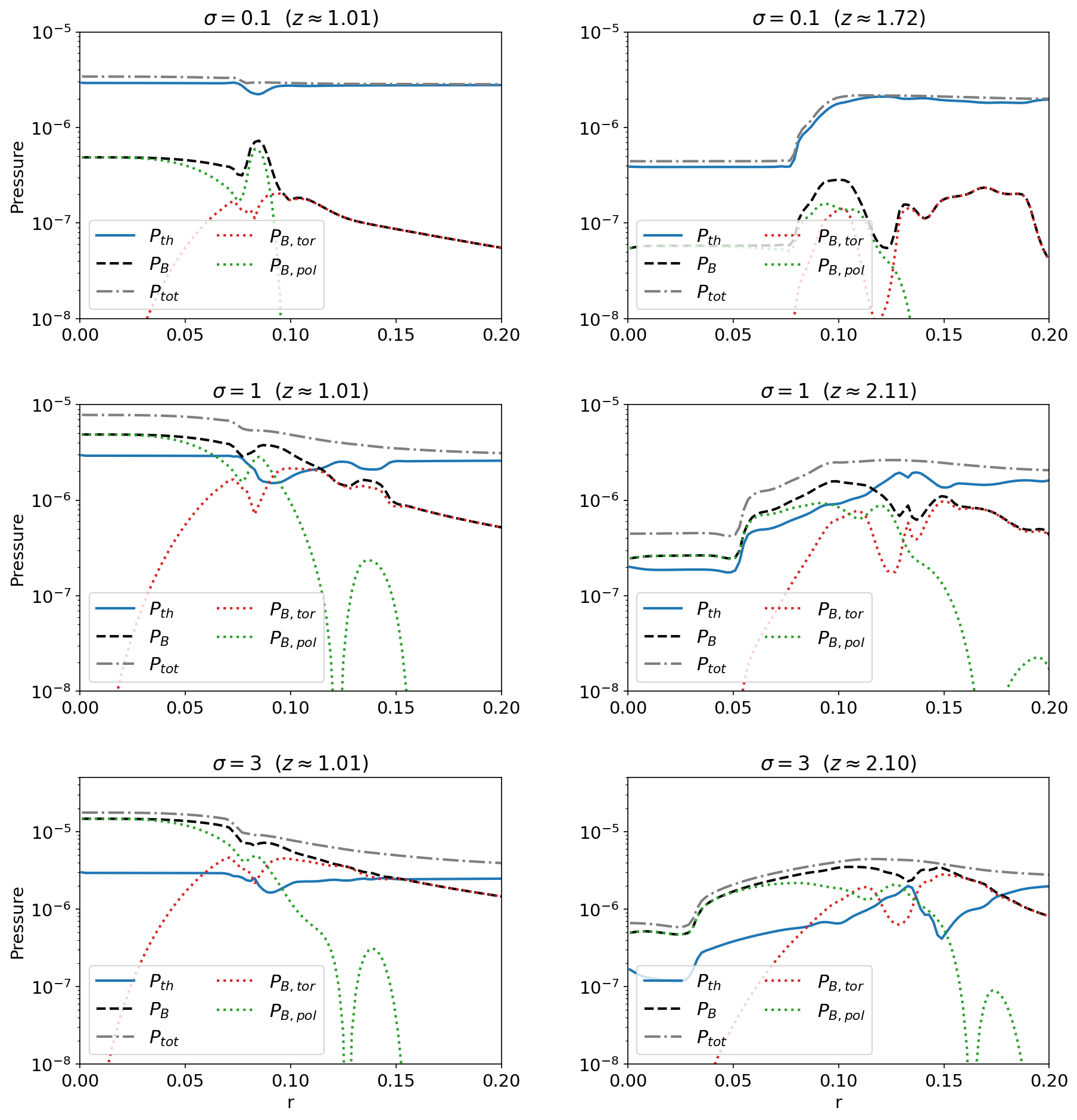}
\caption{
Radial profiles of the thermal pressure, the magnetic pressure and its components, and total pressure for the C-Case model family at $z_0$ and $ z_{\max}$. } \label{fig:sigma_radial_profiles}
\end{center}
\end{figure}
We emphasize that for our default configuration, the presence of a non-negligible magnetization ($\sigma > 0.1$) introduces two major changes; namely, a modification of the shock strength and a systematic decrease in $z_r$. Note that the detail of the shock modification is dependent on the magnetic pitch parameter, as we detail in Sect. 5.2 (in agreement with \citealt{Mizuno2015}).

\subsection{Radial structure and confinement}
To understand the precise physical mechanisms driving both the initial jet expansion and the subsequent recollimation, it is instructive to examine the radial distribution of the thermal, poloidal magnetic, toroidal magnetic, and total pressures. Figure~\ref{fig:sigma_radial_profiles} depicts these profiles for the C-Case model family ($\nu=10^{-5}$, $P_{\rm ratio} \sim 1$) at the injection base ($z_0 \approx 1.01$) and at the downstream location of maximum radial expansion ($z_{\max}$).
At the injection base ($z_0$, left column),  for the weakly magnetized case ($\sigma=0.1$), the core is entirely gas-pressure dominated, with $P_{\rm th}$ remaining uniform out to the boundary layer ($r \approx 0.08$). The pressure balance across all magnetizations supports outward expansion. As the initial magnetization increases to $\sigma=1$ and $\sigma=3$, the core pressure is dominated by the poloidal magnetic component ($P_{\rm B,pol}$), while the toroidal component ($P_{\rm B,tor}$) remains negligible near the axis and only peaks weakly near the jet edge. Because there is no strong, inward-directed total pressure gradient or significant azimuthal hoop stress to balance the internal core pressure at $z_0$, the jet undergoes immediate, rapid lateral expansion.
This structural layout changes radically as the flow moves downstream to the point of maximum expansion ($z_{\max}$, right column), where the external medium effectively confines the jet's boundary. At $z_{\max}$, the profiles show:
(a) For low magnetization ($\sigma = 0.1$, top right), the jet remains hydrodynamically dominated ($P_{\rm th} \gg P_{\rm B}$). Confinement is achieved via a localized thermal pressure barrier; $P_{\rm th}$ drops slightly inside the core due to expansion but matches the ambient pressure at the boundary layer ($r \approx 0.08 - 0.10$), forcing the streamlines to bend back inward. (b) For moderate to high magnetization ($\sigma = 1$ to $\sigma = 3$, middle and bottom right), as $\sigma$ increases, the primary confining shifts from thermal gas pressure to magnetic pressure. At the jet boundary, a steep, inward-directed total magnetic pressure gradient ($P_{\rm B}$) develops, driven by the compressed toroidal field component ($P_{\rm B,tor}$). For $\sigma=3$, the thermal pressure becomes negligible, and the total pressure profile ($P_{\rm tot}$; dash-dotted gray line) is almost entirely dictated by this magnetic pressure wall. This behavior demonstrates that while the lack of an external pressure gradient at the base triggers free expansion, it is the downstream accumulation and compression of the magnetic field at the jet interface that form a powerful confining boundary, driving a faster, more compact recollimation process as the initial magnetization increases.

\subsection{Recollimation shock geometry}
\begin{figure}[htbp] \begin{center} \includegraphics[width=\linewidth]{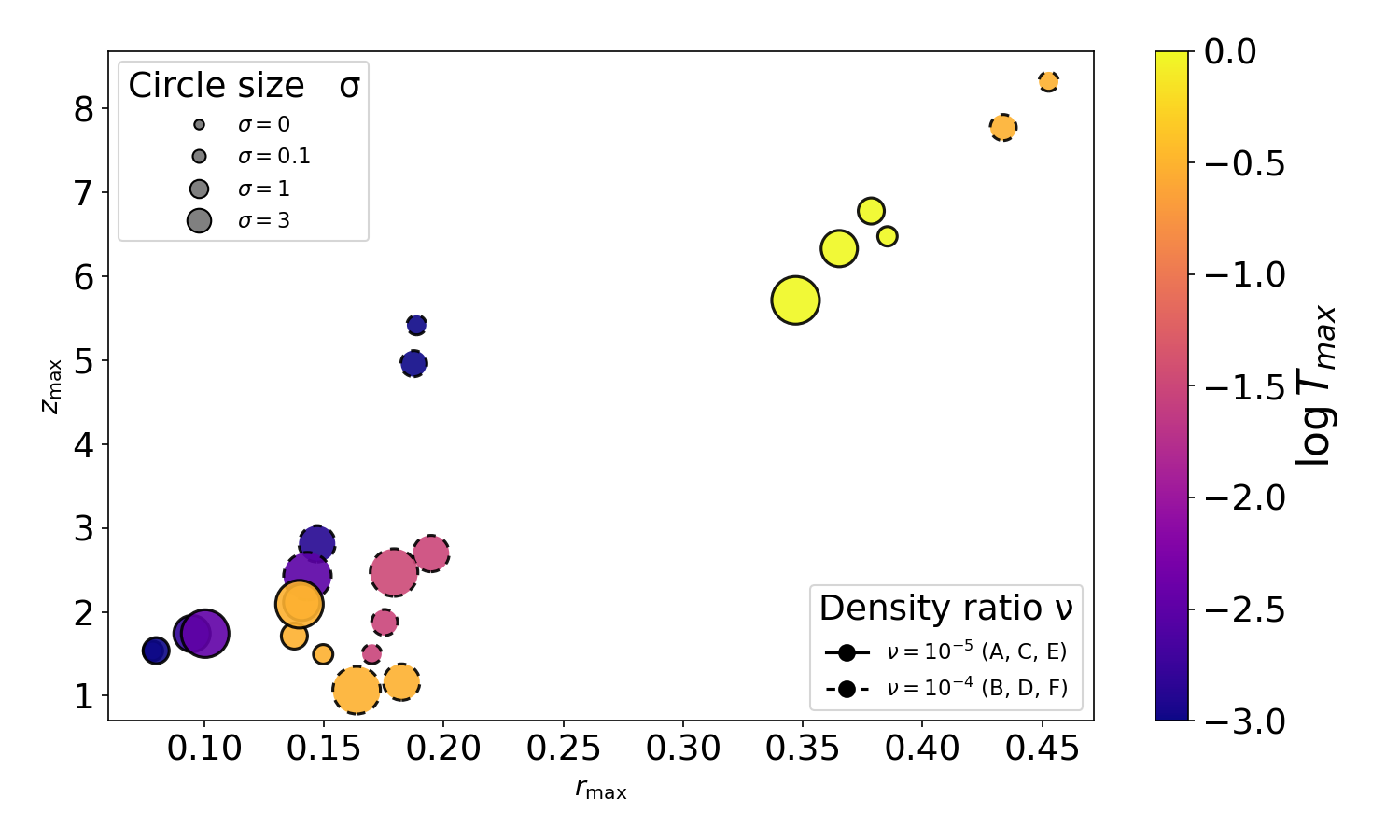} \caption{Global geometrical properties of the first recollimation region ($z_{\max}$ vs $r_{\max}$). The size of the data points corresponds to $\sigma$, and the line style indicates $\nu$. The color map represents $\log(T_{max})$. \label{fig:2D_summary_}} \end{center} \end{figure}

Figure~\ref{fig:2D_summary_} summarizes the global geometry of the first recollimation shock for all models, plotting the axial distance ($z_{\max}$) against the radial extent ($R_{\max}$).
The results clearly validate the trends:
(a) {Magnetization effect ($\sigma$).} For both light and heavy jets, increasing $\sigma$ (larger circles) generally results in smaller $r_{\max}$ and slightly smaller $z_{\max}$. 

    (b) {High thermal domination.} The points with the highest $z_{\max}$ and $r_{\max}$ correspond to models with low $\sigma$ and a high local temperature (e.g., Run E). These are jets dominated by kinetic and thermal energy.
    (c) {Heavy jet confinement.} The dashed-outline points (Runs B and D) generally cluster at higher $z_{\max}$ values compared to their low-density counterparts.

\subsection{Empirical scaling of the recollimation distance}

\begin{figure}[htbp]
\begin{center}
   \includegraphics[width=\linewidth]{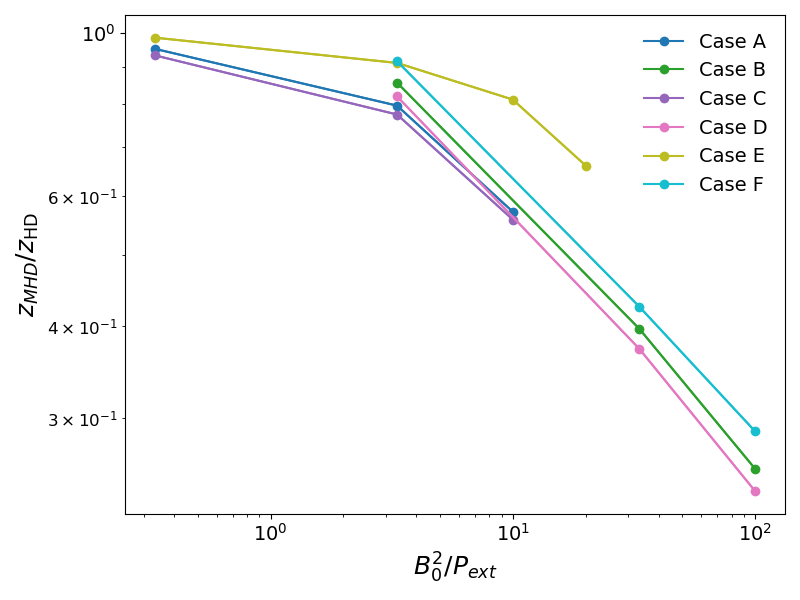}
   \caption{Ratio of the first recollimation shock distance in the MHD case to the purely hydrodynamic case ($z_{MHD}/z_{HD}$) as a function of the initial properties, $B_0^2/P_{ext}$. Note the log-log scale. After an initial plateau at low magnetization, all simulation families (A--F) exhibit a transition to a magnetically dominated regime, converging on a uniform power-law decline with a slope of approximately $-0.33$.}
   \label{fig:z_ratio_sigma}
\end{center}
\end{figure}

To further quantify the impact of the magnetic field on the recollimation geometry, we analyzed the fractional change in the shock position. Figure~\ref{fig:z_ratio_sigma} displays the ratio of the magnetized recollimation distance to the purely hydrodynamic recollimation distance ($z_{MHD}/z_{HD}$) as a function of the injection values ($B_0^2/P_{ext}$).  Unlike the other cases, Case E contains four data points because the run was extended for $\sigma=6$. Cases A and C could not be extended similarly due to numerical problems introduced at very low plasma $\beta$ or high magnetization, as is evident in the density map in Fig. \ref{fig:density_structure_all}.

The plot reveals a striking behavior across the diverse simulation families. At low magnetizations, the recollimation distance remains largely unaffected, with the ratio plateauing near unity. However, as the magnetic field strength increases, a distinct turnover occurs, marking the transition into a magnetically confined regime. Beyond this turnover point, all simulation families within the explored parameter space converge on the same power-law trend, regardless of their density contrasts ($\nu$) or pressure ratios ($P_{\rm ratio}$). In this magnetically dominated state, the curves exhibit a constant logarithmic slope of approximately $-1/3$. This indicates a power-law scaling:
\begin{equation}
z_{MHD}/z_{HD} \propto (B_0^2/P_{ext})^{-1/3}, 
\end{equation}demonstrating that once the magnetic pressure gradient overcomes the thermal pressure gradient, the spatial compression of the recollimation zone follows a predictable empirical scaling law within the explored parameter space.

\section{Role of the magnetic structure}
\label{sec:dynamics}

In this section we analyze how the magnetic field structure, specifically the magnetic pitch parameter, $\alpha$ (Appendix \ref{app:mag_init}), impacts the jet's global morphology, confinement, and internal structure. To cleanly isolate the dynamic effects of the magnetic field topology and evaluate the distinct contributions of the toroidal and poloidal components, we choose to decouple the initial thermal pressure from the magnetic field configuration, by not imposing the magnetothermal equilibrium at the base of the jet. By avoiding a strict setup in which the gas pressure is inherently dependent on the underlying magnetic profile, we ensure that any observed changes in flow behavior are driven solely by variations in the magnetic structure itself.

\subsection{Impact of magnetic pitch on flow expansion and confinement}

\begin{figure*}[htbp]
   \centering
   \includegraphics[width=1.0\linewidth]{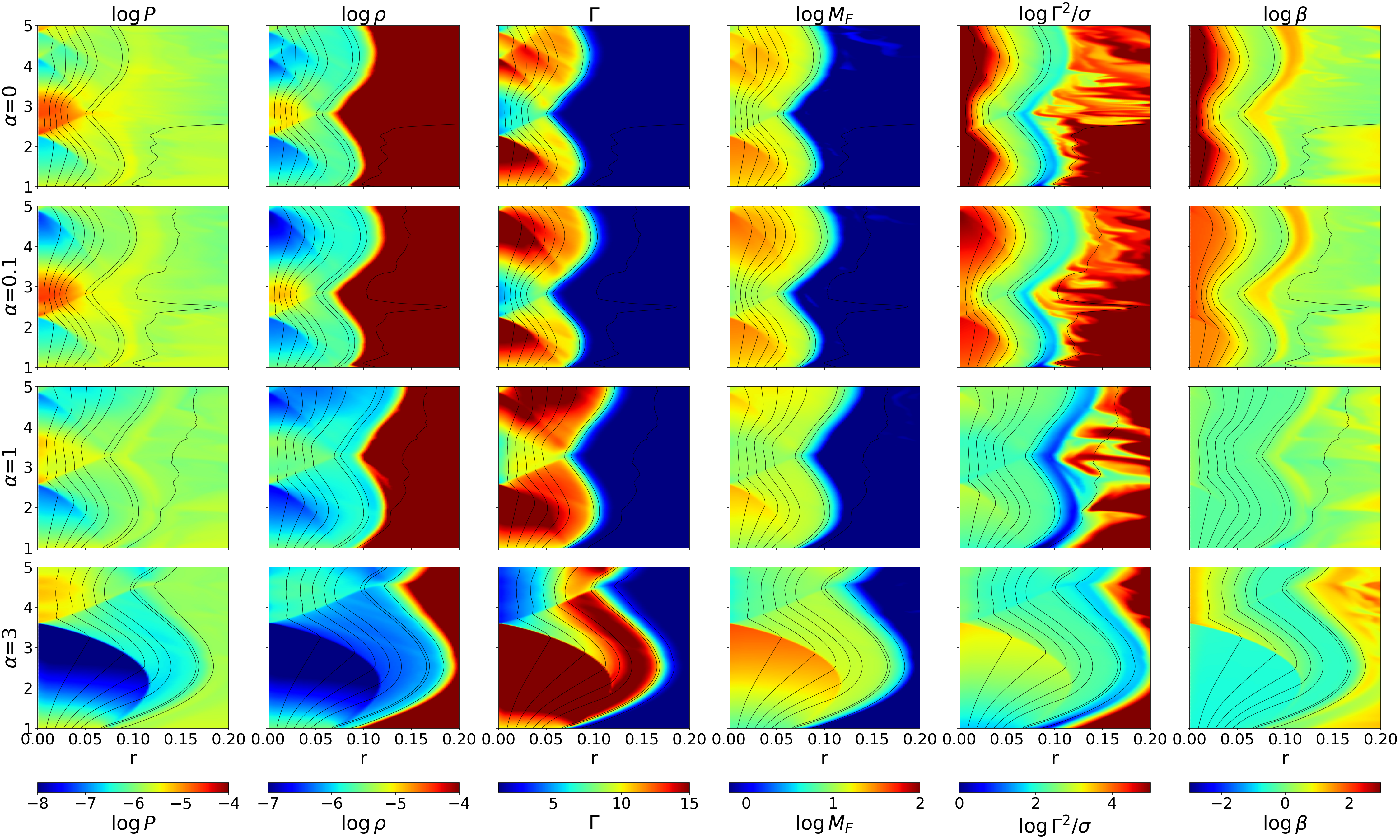}
   \caption{
   Impact of magnetic pitch parameter on jet structure. 
   2D maps in the $(x, z)$ plane showing the steady-state spatial distribution of (from left to right): the logarithm of thermal pressure ($\log P$), logarithm of rest-mass density ($\log \rho$), Lorentz factor ($\Gamma$), logarithm of the fast magnetosonic Mach number ($\log M_F$), stability proxy ($\log (\Gamma^2/\sigma)$), and logarithm of the plasma beta ($\log \beta$).
   The rows correspond to different magnetic pitch parameter values: purely toroidal ($\alpha=0$, top), $\alpha=0.1$, $\alpha=1$, and poloidal-dominated ($\alpha=3$, bottom).
   Black lines indicate streamlines.
   }
   \label{fig:pitch_atlas}
\end{figure*}

\begin{figure*}[htbp]
   \centering
   \includegraphics[width=1.0\linewidth]{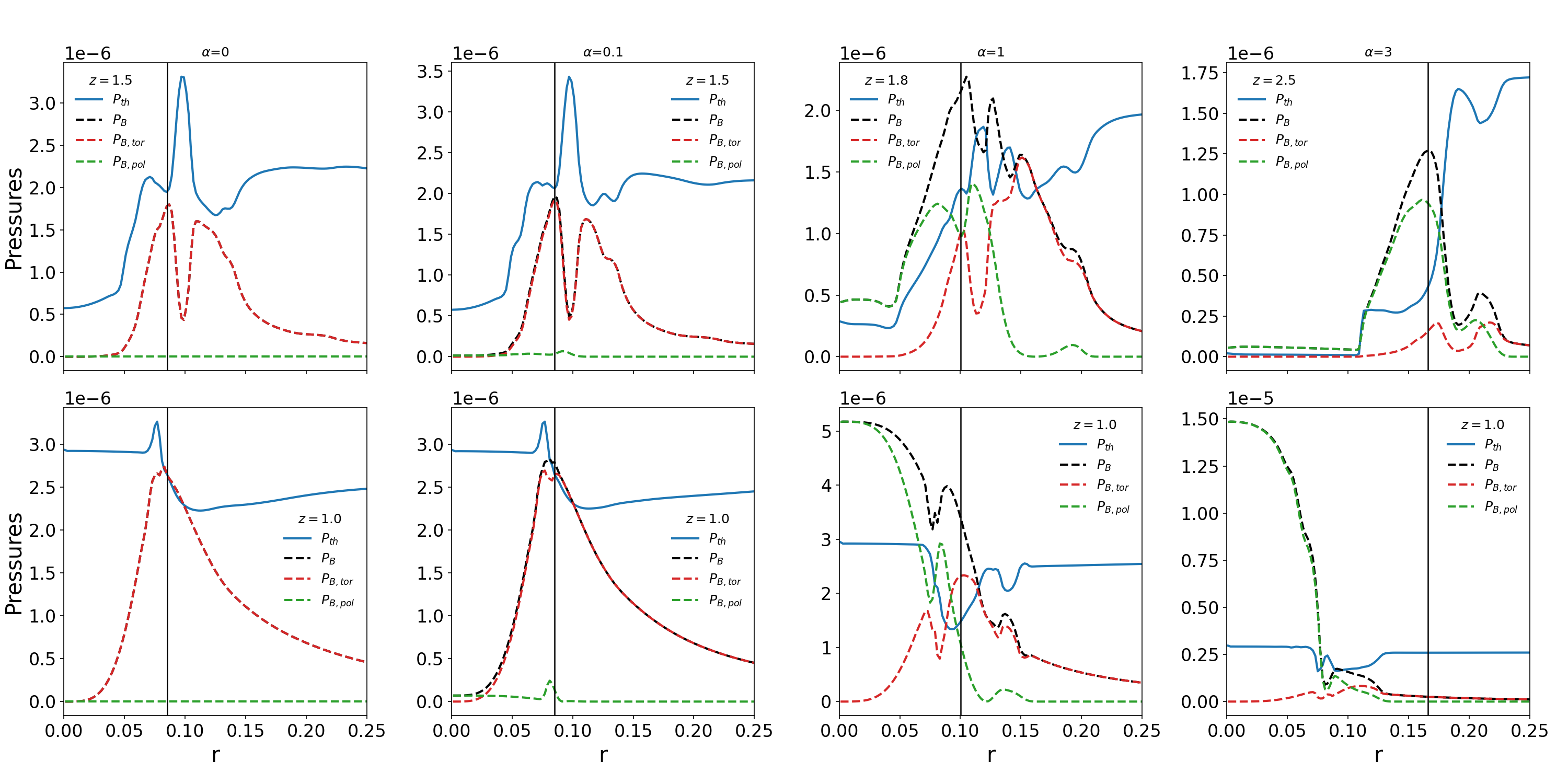}
  \caption{
   Radial pressure balance analysis for increasing magnetic pitch, from purely toroidal ($\alpha=0$, left) to poloidal-dominated ($\alpha=3$, right). \textit{Top panels}: Location of maximum radial expansion ($r_{max}, z_{max}$). \textit{Bottom panels}: Radial pressure balance at $z_0$.
   }
   \label{fig:pitch_forces}
\end{figure*}

Figure \ref{fig:pitch_atlas} provides a detailed atlas of steady-state jet structures for four magnetic pitch parameter values, spanning a purely toroidal configuration ($\alpha=0$) to a poloidal-dominated regime ($\alpha=3$). All cases utilize fixed baseline parameters ($P_{\mathrm{ratio}}=1$, $\sigma=1$, and $\nu = 10^{-5.5} = P_0$). Details regarding the normalization of the magnetic field are provided in Appendix \ref{Ub}.\footnote{Appendix \ref{app:pitch} presents the imposed equilibrium of the radial profiles of thermal and magnetic pressures, in this case, because the thermal pressure and magnetic field are intricately coupled, modifying the pitch also drives up the thermal pressure. This introduces additional complexity ultimately shifting the recollimation point further downstream in the toroidal-dominated case.}
A key observation is the systematic, progressive widening of the jet core and the downstream extension of the recollimation point with increasing pitch. This behavior is directly tied to the  magnetic forces at the base of the flow, Fig. \ref{fig:pitch_forces}):
(a) Toroidal dominance ($\alpha=0 - 0.1$). In the top rows of Fig. \ref{fig:pitch_atlas}, the magnetic field is dominated by the azimuthal component. Close to the base, the initial magnetic configuration produces an inward force related to the dominant magnetic pressure force related to the steep increase in the toroidal field for increasing $r$ (left panel of Fig.\ref{fig:pitch_forces}). This force effectively constrains the lateral expansion of the jet. At higher $z$ (upper panels of Fig.\ref{fig:pitch_forces}), the combined action of inward-directed thermal and magnetic pressure forces effectively arrests the jet’s initial expansion, resulting in the characteristic oscillatory morphology with short-wavelength recollimation shocks ($z \approx 2.2$). 
{(b) Poloidal dominance ($\alpha=3$).} In the bottom row of Fig.\ref{fig:pitch_atlas} , the internal dynamics is inverted. In this configuration, the dominant internal axial field component produces a strong outward-directed force (right panel of Fig. \ref{fig:pitch_forces}). As a result, the jet undergoes a prolonged, gradual expansion driven from within, producing a wider, smoother flow where the first recollimation shock occurs significantly farther downstream ($z \approx 3.5$).

Building on the results for Case B4 shown in Fig.~\ref{fig:z_ratio_sigma}, we investigated the impact of poloidal field dominance by increasing the parameter $\alpha$ from 1 to 3 and 100. We observed that the $z_{MHD}/z_{HD}$ ratio increases from 0.25 to 0.70 and eventually reaches 1.05 for $\alpha=100$. This trend demonstrates that as the poloidal component becomes dominant, its internal pressure actively promotes jet expansion. This shifts the recollimation point downstream, to distances that can even exceed those found in the purely hydrodynamic case.

\subsection{Shock structure and strength}
\label{sec:shock_structure}

\begin{figure*}[htbp]
   \centering
   \includegraphics[width=1.0\linewidth]{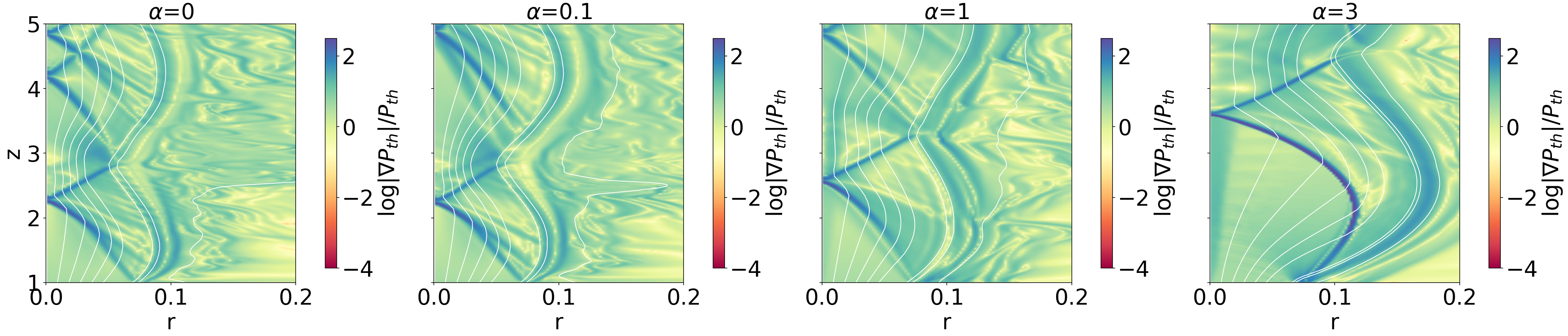}
   \caption{
 Normalized pressure gradient ($|\nabla P_{th}|/P_{th}$). The white lines are the streamlines.
   }
   \label{fig:pitch_gradients}
\end{figure*}

The variation in the magnetic pitch parameter, $\alpha$, fundamentally alters the resulting shock structure of the jet, Fig. \ref{fig:pitch_gradients}. The magnetic field configuration dictates not only the flow acceleration profile but also the plasma's intrinsic compressibility at shock fronts. 

\begin{figure}[htbp]
    \centering
    \includegraphics[width=\linewidth]{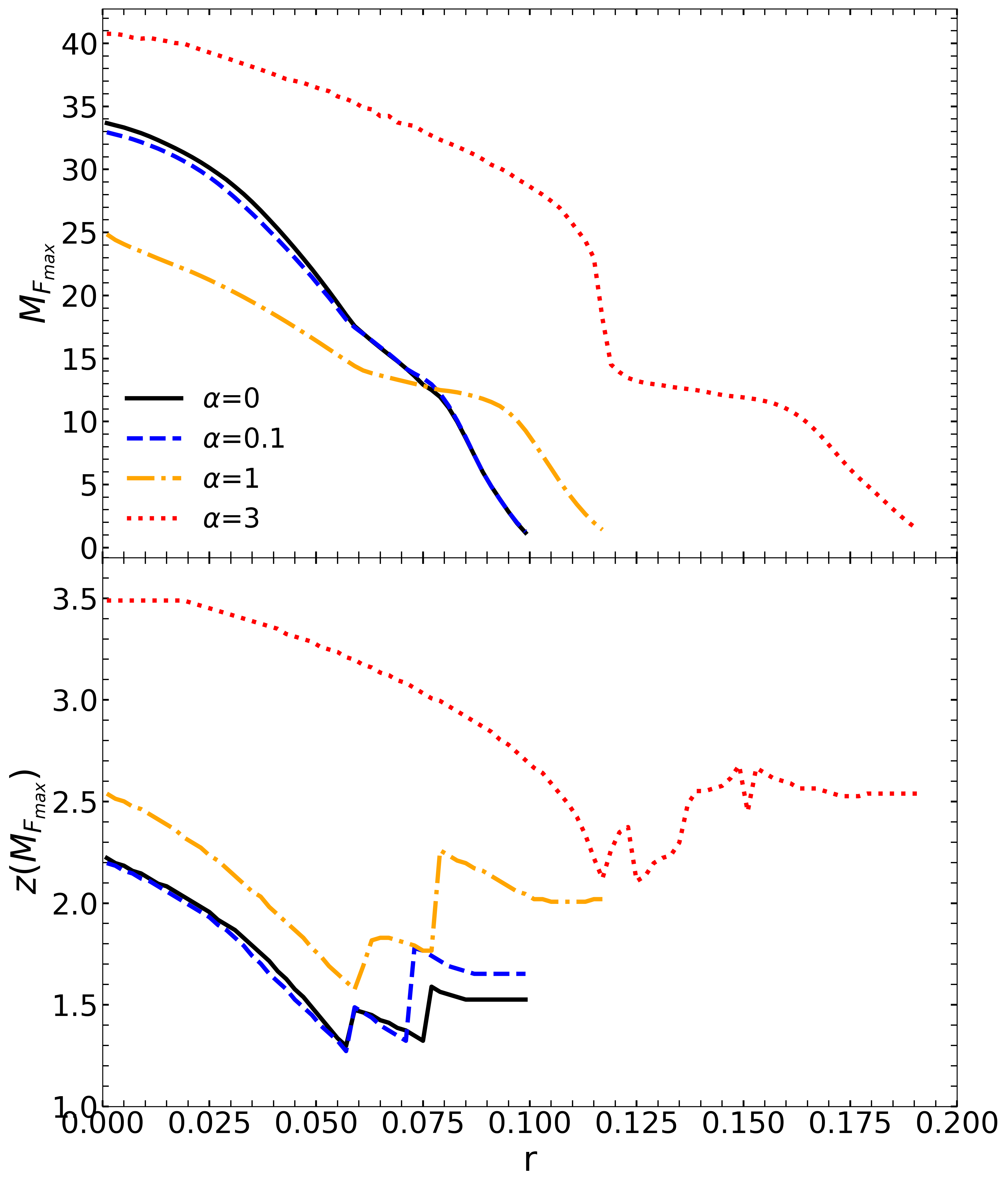} 
    \caption{Radial evolution of the maximum fast magnetosonic Mach number and its corresponding vertical height. \textit{Top}: Maximum value of $M_{F_{\max}}$ evaluated over the domain $z \in [1.0, 3.5]$ as a function of the radial distance $r$. \textit{Bottom}: $z$ coordinate at which this maximum is located, effectively tracing the trajectory of the primary shock front or wave structure. The different curves denote magnetic pitch parameters of $\alpha=0$ (solid black), $\alpha=0.1$ (dashed blue), $\alpha=1$ (dash-dotted orange), and $\alpha=3$ (dotted red). The profiles are explicitly truncated where  $M_F < 1$, indicating the spatial point where the primary structure attenuates into the ambient medium.}
    \label{fig:mf_max_evolution}
\end{figure}

In configurations dominated by a toroidal magnetic field ($\alpha=0$ and $0.1$), the magnetic field lines are oriented primarily parallel to the recollimation shock fronts (perpendicular to the shock normal). A magnetic field component parallel to the shock surface introduces strong transverse magnetic pressure. This added magnetic pressure resists compression, reducing the plasma's compressibility and thereby naturally weakening the resulting shock. Conversely, as the poloidal magnetic field component becomes dominant ($\alpha=3$), the jet dynamics change significantly. The strong poloidal magnetic pressure contributes directly to the outward expansion and longitudinal acceleration of the jet. Consequently, the flow in this regime becomes heavily dominated by kinetic energy flux. Because the kinetic energy is substantially higher than in toroidal-dominated cases, the flow has much greater inertia, leading to significantly stronger shocks when the jet core interacts with the surrounding boundaries.
Interestingly, the intermediate case ($\alpha=1$) deviates from this general trend. Because this configuration is in a state of magneto-thermal equilibrium, the flow undergoes a much smoother initial expansion without immediately triggering the violent shock discontinuities seen in the other configurations. Rather than following the extreme behaviors of the purely toroidal or highly poloidal regimes, the $\alpha=1$ jet maintains a balanced, transitional state before eventually experiencing boundary interactions.
To rigorously quantify the strength and spatial persistence of these wave structures, we analyzed the fast magnetosonic Mach number, $M_F$. Because the shock fronts' vertical positions vary across different pitch configurations, radial profiles at fixed heights are insufficient. Instead, we extracted the absolute maximum value of $M_F$ along the vertical $z$ axis for every radial distance, $r$. 
As is illustrated in Fig. \ref{fig:mf_max_evolution}, this diagnostic clearly demonstrates the enhanced shock strength in the poloidal regime. The highly poloidal case ($\alpha=3$) maintains a strongly bast-magnetosonic core (reaching $M_F > 35$) that persists much further outward in radius. In contrast, the toroidal-dominated flows ($\alpha=0$) possess weaker initial Mach numbers and experience a rapid decline, rapidly dissipating and transitioning into the sub-fast regime ($M_F < 1$) at significantly smaller radii. 

\subsection{Synthetic synchrotron maps}

\label{sec:synchrotron_morphology}

To evaluate how the structural differences discussed above manifest observationally, we constructed synthetic synchrotron emission maps following the approach of \cite{BT2018}. The local, bolometric synchrotron emissivity is assumed to scale with the thermal pressure ($P_{\mathrm{th}}$), which serves as a proxy for the relativistic electron energy density, and with the local magnetic field pressure ($B$). Accounting for relativistic  Doppler boosting for a jet viewed at an observation angle, $\theta$, with respect to the axis, the emissivity in the observer’s frame ($\varepsilon_{\mathrm{obs}}$) is given by
\begin{equation}
\varepsilon_{\mathrm{obs}} = \varepsilon_{\mathrm{com}} \delta^{3} \propto P_{\mathrm{th}} P_B \delta^{3},
\end{equation}
where $\delta = [\gamma(1 - {\boldsymbol{\beta \cdot}}\mathbf{n})]^{-1}$ is the relativistic Doppler factor. By linking $\varepsilon_{\mathrm{obs}}$ directly to the underlying MHD variables, we map the visual signature of the internal flow across different magnetic configurations.

\begin{figure*}[htbp]
    \centering
    \includegraphics[width=\textwidth]{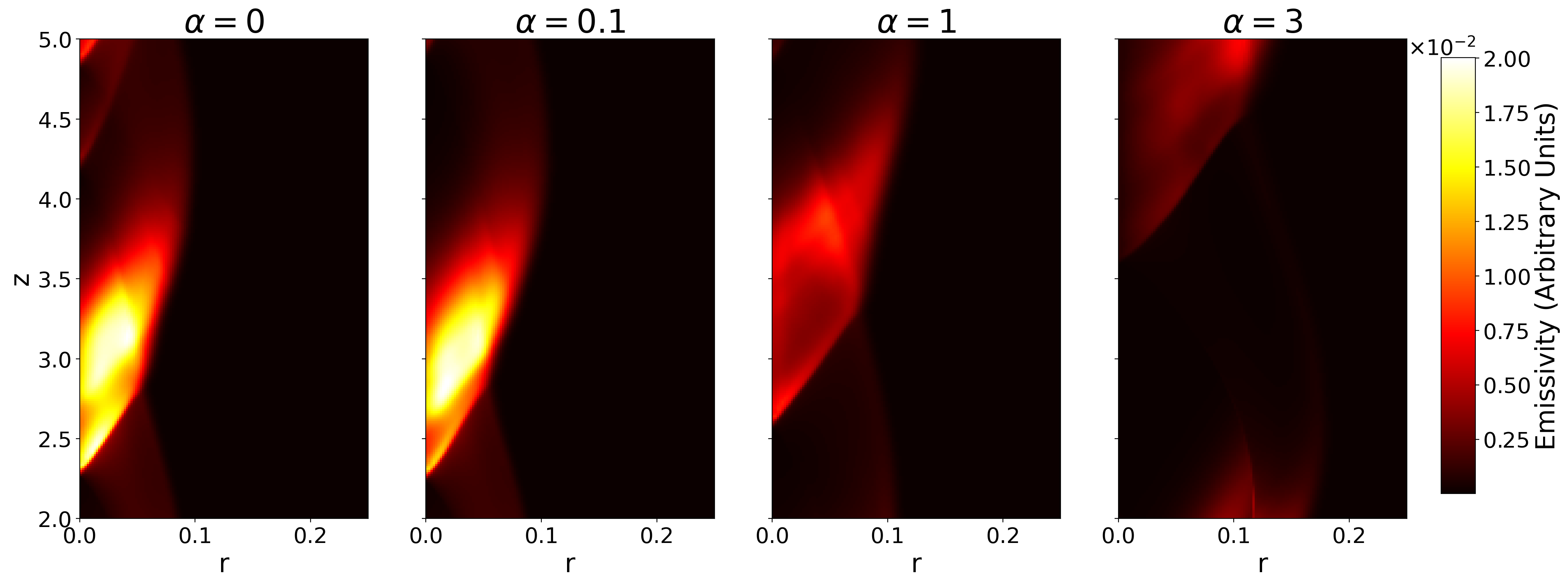}
    \caption{2D synthetic synchrotron emission maps ($\varepsilon_{\mathrm{obs}}$) for a viewing angle of $\theta = 5^{\circ}$, comparing different magnetic pitch parameters profiles ($\alpha = 0, 0.1, 1, 3$). The maps focus on the structural layout surrounding the primary recollimation shocks. }
    \label{fig:synchrotron}
\end{figure*}

Figure \ref{fig:synchrotron} displays the resulting observer-frame emissivity maps focused around the primary recollimation and reflection shock region ($z \in [2.0, 5.0]$) for an observation angle of $\theta = 5^{\circ}$. The structural transition discussed in Sect. \ref{sec:shock_structure} is clearly reflected in the intensity profiles. In the low-pitch parameter regimes ($\text{pitch} = 0$ and $0.1$), the frequent, thin shock filaments identified via pressure gradients materialize as localized, highly boosted brightness enhancements. The strong reflection  shock induces a sharp spike in both thermal pressure and magnetic field compression, concentrating the emission into a bright, narrow knot near $z \approx 2.8$ with a tight radial profile ($r < 0.05$). Conversely, as the magnetic pitch parameter increases ($\text{pitch} = 1$ and $3$), the enhanced expansion provided by the poloidal field contributes to lowering the thermal and magnetic pressures in the downstream regions of both the recollimation and the reflection shocks, resulting in a weaker and more diffuse emission.

It is important to note a physical limitation of our emissivity proxy. The synthetic maps rely on the macroscopic thermal and magnetic pressures to trace emission; however, the recollimation shocks driving this compression are inherently oblique. In relativistic environments, such oblique shocks are generally superluminal, meaning the intersection point of the magnetic field lines and the shock front moves faster than the speed of light. This geometry prevents particles from escaping upstream, strongly suppressing standard diffusive shock acceleration. As discussed by \cite{sciaccaluga25}, efficient particle acceleration at these specific sites ultimately requires the presence of significant magnetic turbulence (see references therein). Turbulence can induce local magnetic field line wandering, temporarily restoring subluminal conditions and allowing cross-field diffusion to sustain the acceleration process. Therefore, while our maps highlight the macroscopic regions most favorable for enhanced emission, the exact microscopic efficiency of these knots heavily relies on the turbulence generated within the shocked layer.

\section{Stability analysis: Centrifugal instability}
\label{sec:cfi}

The CFI is a local, curvature-driven mode that can develop in jets when the destabilizing effect of motion along bent streamlines overcomes the stabilizing tension of the toroidal magnetic field. 

\subsection{Mapping the instability}
\begin{figure}[htbp] 
\begin{center}
\includegraphics[width=\linewidth]{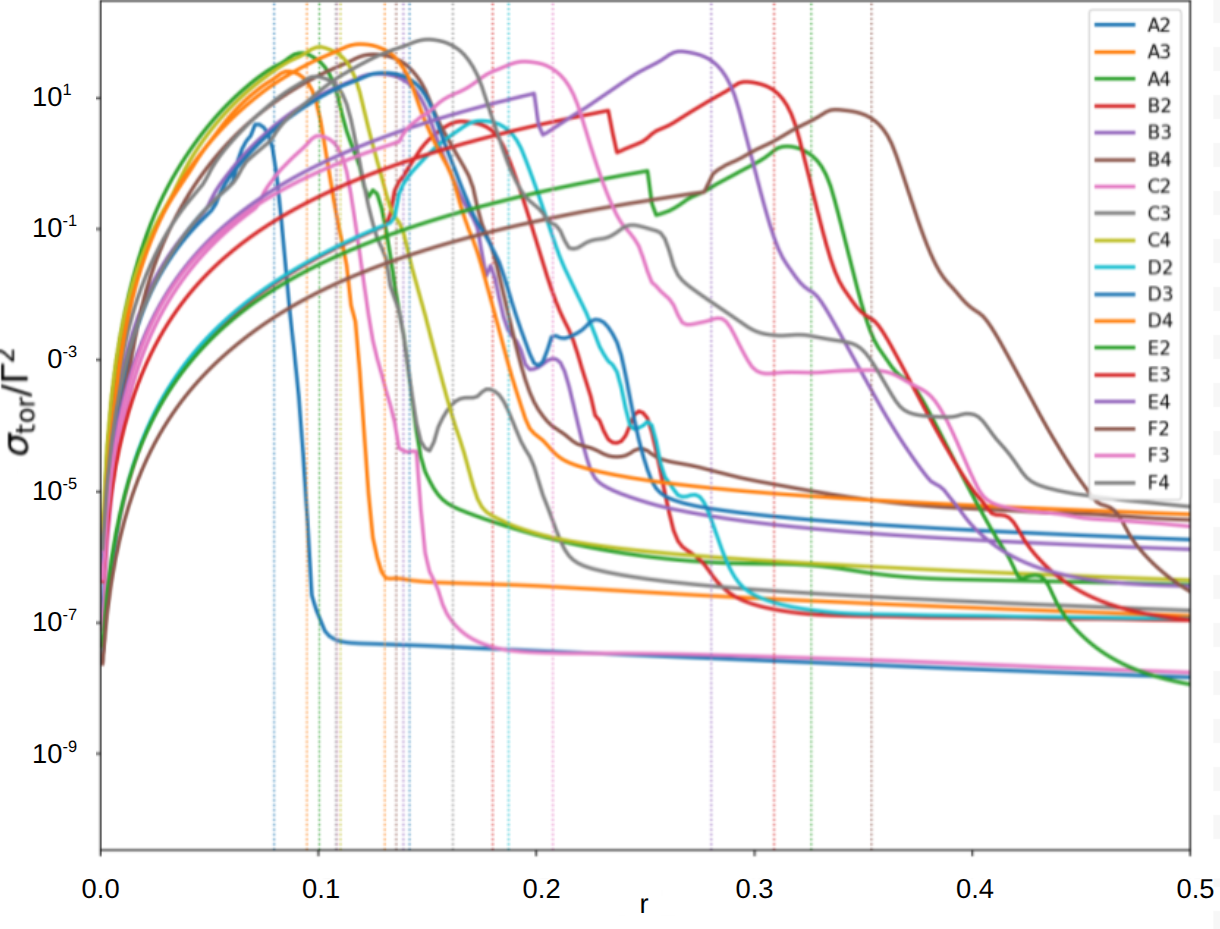} \caption{ Radial profiles of the normalized toroidal magnetization, ${\sigma_{\mathrm{tor}}/\Gamma^2}$, as a function of the radial coordinate, ${r}$, for all simulated cases that correspond to the $z_{max} $  of the  streamline that starts from the position $r_{\text{streamline}} = 0.075$. \label{fig:2Dcfi}}
\end{center}
\end{figure}

Figure \ref{fig:2Dcfi} presents the radial profiles of the diagnostic ratio $\sigma_{\mathrm{tor}}/\Gamma^2$ for all cases at $z_{max}$ of a specific magnetic streamline,  $r_{stremaline}=0.075$. The profiles show complex structures with sharp drops, often reaching minimum values near the recollimation shocks. A minimum value of $\sigma_{\mathrm{tor}}/\Gamma^2$ indicates the location most prone to CFI.
Table \ref{tab:sigma_curvature_compact} summarizes the values. Generally, cases with lower magnetization (A and C) show a trend toward lower $\sigma_{\mathrm{tor}}/\Gamma^2$ values, suggesting higher susceptibility.

\subsection{Linear growth rates}
Although linear analyses of idealized cylindrical shells offer fundamental insights into the CFI mechanism \citep[e.g.,][]{Komissarov2019}, fully 3D RMHD simulations indicate that instability development is highly sensitive to the precise transverse gradients formed during recollimation \citep[see][]{Boula25,Musso2024}. Therefore, it is necessary to perform linear calculations directly on the evolved simulation data to accurately capture the structure of the magnetic and thermal profiles near the non-smooth shear layer.

\begin{figure}[htbp]
\begin{center}
\includegraphics[width=\linewidth]{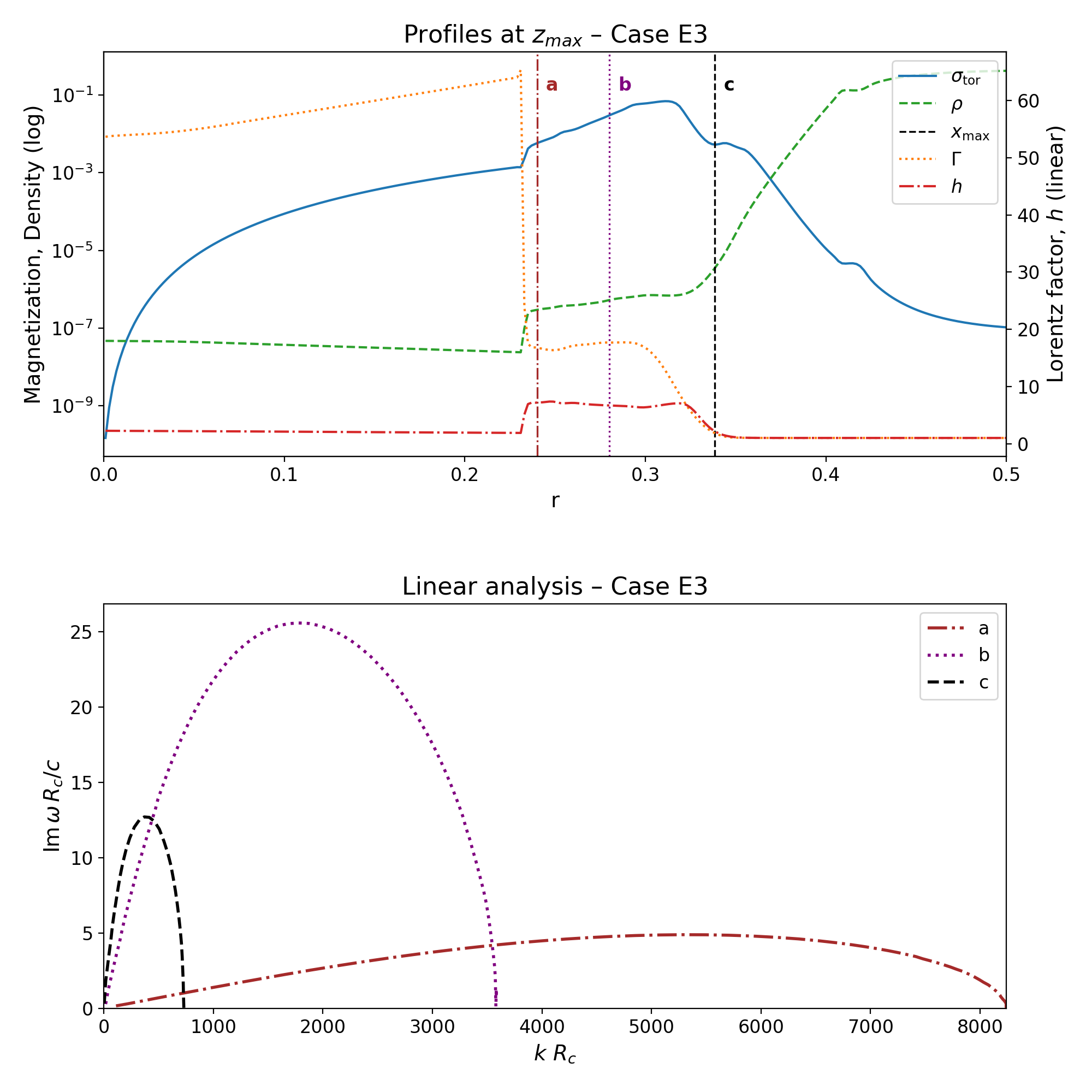}
\caption{Radial profiles and linear analysis results for the Case E3 jet near its recollimation point. \textit{Top}: Radial profiles of key quantities used as input for the linear analysis: normalized toroidal magnetization ($\sigma_{\text{tor}}$), rest mass density ($\rho$), Lorentz factor ($\Gamma$), and specific enthalpy ($h$). The vertical lines marked a, b, and c indicate the specific radial positions at $x_{\mathrm{max}}$ where the linear analysis was performed. \textit{Bottom}: Growth rate ($\Im \omega$) spectrum of the CFI as a function of the normalized wavenumber, $k R_c$, for the three positions indicated in the top panel.}\label{fig:CaseE}
\end{center}
\end{figure}

To quantitatively validate the susceptibility indicated by the $\sigma_{\text{tor}}/\Gamma^2$ mapping, a localized linear stability analysis was conducted using evolved profiles extracted directly from the fiducial A3 model (see Appendix \ref{sec:linear} for details). Figure \ref{fig:linear} (bottom panel) presents the resulting growth rate spectrum ($\Im \omega$). The unstable region (red) corresponds to the expected growth of the CFI mode.
The profiles exhibit a pronounced sensitivity to the streamline radius (Tab. \ref{tab:sigma_curvature_compact}). This rapid variation reflects the complex and non-smooth structure of the flow near the shear layer, reinforcing the necessity of rigorous local linear analysis to determine where instability may develop.
\begin{figure}[htbp] 
\begin{center}
\includegraphics[width=\linewidth]{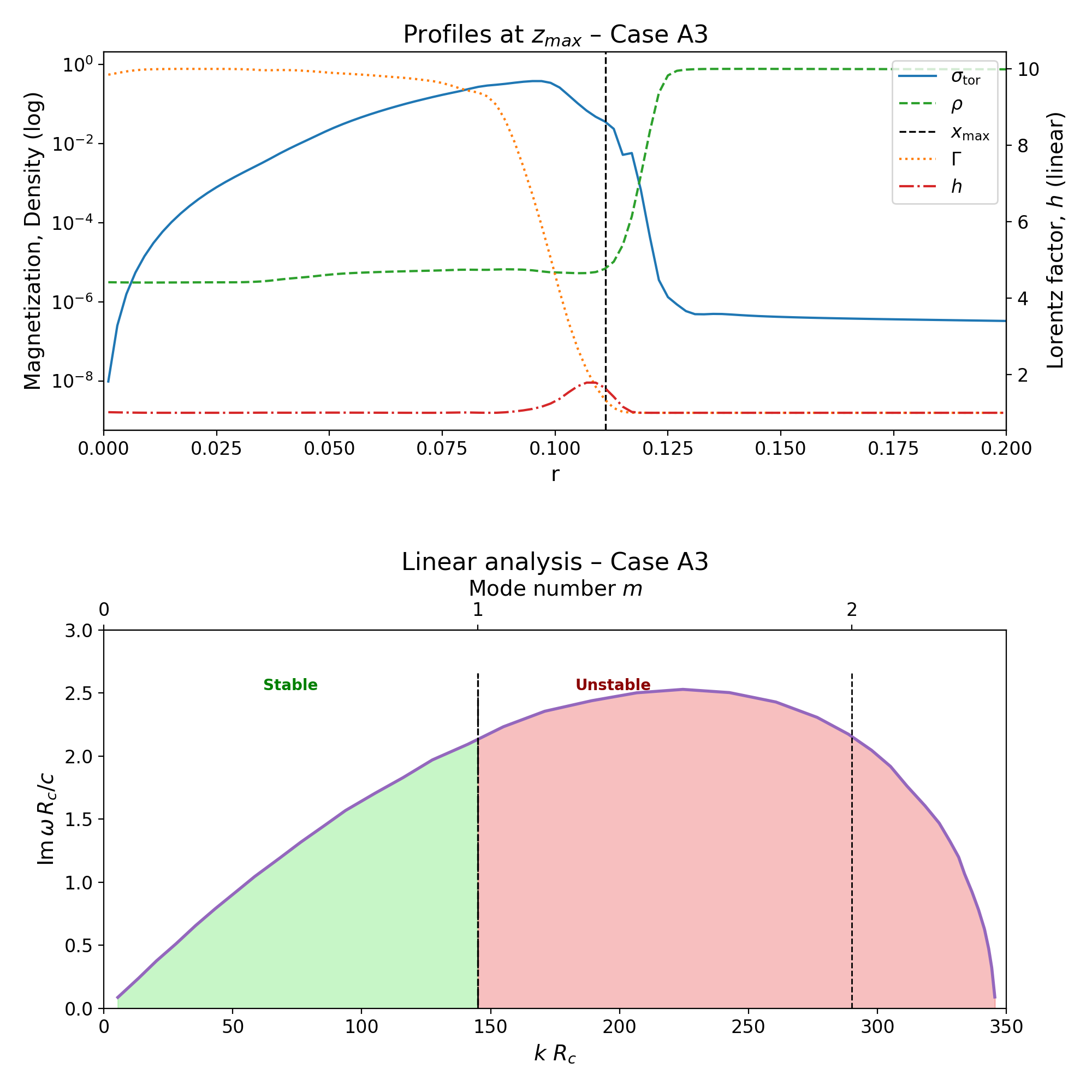} \caption{\textit{Top}: Radial profiles of key quantities for the A3 Case jet near the recollimation point. \textit{Bottom}: Growth rate ($\Im \omega$) spectrum of the CFI as a function of the normalized wavenumber, $k R_c$. \label{fig:linear}}
\end{center}
\end{figure}
The application of this local linear analysis to the 2D simulation data, as shown in Fig. \ref{fig:CaseE}, highlights a critical challenge. The top panel of Fig. \ref{fig:CaseE} presents the radial profiles of key quantities for a specific case (E3) near the recollimation point, where the flow variables are continuously varying and the shock transition is not perfectly sharp. The linear analysis was performed by selecting local conditions at three distinct radial positions ($a$, $b$, and $c$). As is demonstrated in the bottom panel of Fig. \ref{fig:CaseE}, the resulting growth rate ($\Im \omega$) spectrum changes dramatically depending on which position is chosen. This sensitivity underscores the difficulty of obtaining a definitive stability result when the underlying physical profiles are neither sharp nor extended. Such complicated, nonideal profiles suggest that a more sophisticated approach, such as a global eigenmode analysis that accounts for the entire profile shape and boundary conditions, may be necessary for accurate predictions.
 
\section{Discussion}
\label{sec:discussion}
Our steady-state RMHD simulations demonstrate a complex interaction between jet hydrodynamics, magnetic field strength and structure, and ambient confinement that shapes the large-scale dynamics and energetics of AGN outflows. Radial analysis (Fig.~\ref{fig:sigma_radial_profiles}) confirms that the  magnetic field strongly impacts  the jet's global geometry.  
Although dense ambient media ($\nu=10^{-4}$) further compress the flow and increase maximum field strengths, magnetization ($\sigma$) remains the key parameter governing the jet's energetic transition and morphology.

We identify the injection magnetization, $\sigma$, as the primary parameter controlling energy conversion along the jet. As shown in Fig.~\ref{fig:z_ratio_sigma}, increasing $\sigma$ (for a fixed magnetic pitch parameter) results in a monotonic decrease in the position of the first recollimation point, directly reflecting that the stronger outer magnetic pressure gradient efficiently limits the initial free expansion phase and triggers earlier recollimation. For highly magnetized runs ($\sigma \ge 1$), the flow confirms a magnetically dominated regime. Due to the high inertia, the conversion of specific enthalpy into bulk kinetic energy occurs gradually during recompression. This acceleration culminates at the recollimation shock front, where the bulk kinetic energy is ultimately dissipated. Regarding general jet evolution, our results show dynamical trends qualitatively consistent with the findings of \cite{MT21}, highlighting the magnetic field as the central architect of the jet structure despite the differing numerical configurations. This alignment suggests a robust physical behavior in magnetized jets that transcends specific setup variations.

A particularly significant finding, shown in Fig.~\ref{fig:z_ratio_sigma}, is that once the magnetic field dominates jet confinement, marked by a turnover point where all simulated jet families follow nearly identical trajectories. Regardless of ambient density contrast, pressure ratio, or internal thermal state, these curves display a consistent logarithmic slope of about $-0.33$. This scaling, $z_{MHD}/z_{HD} \propto (B_0^2/P_{ext})^{-1/3}$, shows that after the transition to a magnetically dominated regime, spatial compression of the recollimation zone follows a predictable, parameter-independent scaling law. In future work, the role of the external environment will be tested.

Beyond the overall magnetization scale, the internal structural geometry and observational appearance of the recollimation zone are heavily dictated by the magnetic pitch configuration. Our analysis reveals a distinct transition between a shock-dominated morphology at low pitch parameter values ($\alpha \le 0.1$) and a smooth expansion regime at high pitch parameter values ($\alpha = 3$). In the toroidal-dominated case, the enhanced compression drives up local thermal and magnetic energy densities, which manifests as highly boosted, localized synthetic synchrotron emission knots concentrated near the primary reflection shock ($z \approx 2.8$, $R < 0.05$). Conversely, a dominant poloidal component ($\alpha = 3$) contributes strong magnetic pressure that drives significant acceleration, causing the expanding jet core to become heavily dominated by kinetic energy flux. Rather than dropping to subfast values, this high inertia flow maintains a strongly fast magnetosonic state ($M_F \gg 1$, as seen in Fig. \ref{fig:mf_max_evolution}) over an extended spatial domain. Because the plasma is moving too fast to communicate pressure changes upstream, it cannot adjust smoothly to external environmental gradients. Consequently, when this high kinetic energy flow inevitably interacts with the boundary, it generates significantly stronger and more persistent shock discontinuities, pushing the primary recollimation region much further downstream to $z \approx 3.8$ and producing an elongated, diffuse observer-frame emission profile.

Relating these axisymmetric results to observed jet stability requires an understanding of the triggers of non-axisymmetric instabilities. A key conclusion is that the conditions for CFI, which drives 3D dissipation downstream of recollimation shocks \citep{GK18b, Matsumoto2021, Boula25}, are determined by the evolved steady-state flow topology rather than injection parameters. The instability depends on the local radial profiles of the destabilizing term $1/\Gamma^2$ and the stabilizing normalized toroidal magnetization, $\sigma_{\rm tor}$, both of which are formed within the shocked layer. Our analysis offers a diagnostic tool by identifying regions most susceptible to CFI, where streamline curvature is highest and specific magnetization is lowest. Small differences in steady-state geometry can produce either stable, collimated jets (FR1 and FR2 outflows) or the unstable, turbulent jets typical of FR0 jets. Quantifying the CFI diagnostic in recollimation regions of observed sources such as HST-1 in M87, where stationary recollimation structures develop significant downstream instabilities \citep{Walker2018}, provides a valuable tool for predicting source evolution.

The scaling identified here provides useful insights for interpreting observational data, particularly in the context of statistical population studies. Stationary knots in blazar jets are frequently modeled as recollimation shocks \citep{Cohen14, Paraschos2025}. If high-resolution VLBI observations can locate the first stationary knot relative to the core and combine this with environmental density estimates or jet width profiles, our scaling law offers a potential framework to place broad constraints on jet magnetization without requiring extensive, source-specific numerical modeling. The position of the recollimation points may also help refine models of flaring blazars such as Mrk 421, where multiple stationary shocks have been proposed to explain complex X-ray flaring behavior \citep{Hervet2019}. By qualitatively relating the spacing between stationary knots (likely corresponding to our $z_{\max}$) to the magnetization via the $(B_0^2/P_{\rm ext})^{-1/3}$ trend, these results can help guide the initial background profiles used in self-consistent multi-zone radiation modeling \citep[e.g.,][]{BMK26}.

From a methodological perspective, using highly resolved 2D axisymmetric base flows offers a clear advantage for this study, which is not perturbed by early 3D turbulent degradation. While CFI requires 3D geometries to manifest fully, our 2D steady profiles provide a necessary baseline for subsequent stability analyses. Furthermore, the localized linear stability diagnostic is sensitive to the exact radial coordinate chosen within the continuously varying shocked shear layer. This spatial dependence demonstrates the intrinsic limitations of applying local algebraic criteria to non-smooth RMHD boundary layers, underscoring that a comprehensive global eigenmode analysis or fully 3D numerical simulations represent the definitive next steps for quantifying the absolute growth and saturation of these instabilities.

\section{Conclusions}
\label{sec:conclusions}
We have presented the results of a systematic survey of 2D axisymmetric RMHD simulations. Our key findings are:
\begin{itemize}
    \item {Recollimation scaling:} Beyond a threshold magnetic strength, the recollimation distance transitions to a magnetically dominated regime, where all simulated environments within the explored parameter space follow the empirical scaling, $z_{MHD}/z_{HD} \propto (B_0^2/P_{ext})^{-1/3}$. 
    \item {Magnetic pressure and confinement:} The jet's global geometry is primarily confined by an inward-directed magnetic pressure gradient that builds up at the jet boundary. The recollimation distance decreases monotonically with increasing $\sigma$, as this magnetic pressure layer actively limits radial expansion.
    \item {Transition to magnetic dominance:} As magnetization increases ($\sigma \gtrsim 1$), the shock becomes weaker.
   \item{Role of the magnetic field topology:} Contrary to simple geometric expectations, our results demonstrate that the structure, location, and emissivity of the recollimation zone depend on the magnetic configuration in a highly non-trivial manner. Analysis of the magnetic pitch parameter reveals that a dominant toroidal component concentrates energy into highly boosted, localized emission knots near the primary reflection shock. Conversely, an increasing poloidal component introduces complex pressure alterations that extend the expansion phase and shift the recollimation zone downstream, resulting in a significantly more elongated and diffuse emission profile.
    \item {Stability is a localized, evolved property:} The conditions favorable for CFI are governed by the evolved geometric structure. We identify regions predicted by linear theory to be susceptible to CFI, where the combination of high streamline curvature and low $\sigma_{\text{tor}}/\Gamma^2$ satisfies the linear stability criterion.
\end{itemize}

\begin{acknowledgements}
We thank the anonymous referee for their useful comments. We gratefully acknowledge financial support by INAF Theory Grant 2022, 2024 (PI F. Tavecchio).   We acknowledge support by CINECA, through ISCRA and Accordo Quadro INAF-CINECA, and by PLEIADI, INAF – USC VIII, for the availability of HPC resources (PI S. Boula). P.C. was supported by NASA IXPE  Guest Observer grants 80NSSC24K1177 and 80NSSC24K1764. We have used the following Python libraries: Numpy \citep{numpy}, Matplotlib \citep{matplotlib}, Scipy \citep{scipy}, and PyPluto \cite{Pypluto25}.
\end{acknowledgements}

\bibliographystyle{aa}
\bibliography{refs} 

\begin{appendix}
\section{Numerical setup}\label{app:3dres}
Similar to \cite{Boula25} we perform 2D axisymmetric simulations to identify steady-state solutions featuring a clear recollimation structure. 
Specifically, a relativistic conical jet with an opening angle \( \theta_j \) is launched into the computational domain at a distance \( z_0 \) from the cone's apex, embedded within a confining ambient medium. We extend the setup by incorporating magnetic fields into the simulations.

The system of equations in conservation form reads Eq. \ref{eq:1}.
The set of primitive variables is given by the density in the rest frame $ \rho $, the thermal pressure $ p $, the three-velocity $\mathbf{u}$ in the laboratory frame and the three-vector magnetic field $\mathbf{B}$ in the laboratory frame.
Here, $\Gamma$ denotes the Lorentz factor, $\mathbf{f}_g$ is an external force density in the lab frame, and $\mathbf{I}$ represents the $3 \times 3$ identity tensor. The conserved variables include the magnetic field $ \mathbf{B} $, as well as the lab-frame mass, momentum, and energy densities, defined as

\begin{equation}
\begin{aligned}
D &= \Gamma \rho, \\
\mathbf{m} &= w_t \Gamma^2 \mathbf{u} - b_0 \mathbf{b}, \\
E_t &= w_t \Gamma^2 - b_0b_0 - p_t,
\end{aligned}
\qquad \text{with} \qquad
\begin{aligned}
b_0 &= \Gamma\, \mathbf{u} \cdot \mathbf{B}, \\
\mathbf{b} &= \frac{\mathbf{B}}{\Gamma} + \Gamma\, (\mathbf{u} \cdot \mathbf{B})\, \mathbf{u}, \\
w_t &= \rho h + \frac{B^2}{\Gamma^2} + (\mathbf{u} \cdot \mathbf{B})^2, \\
p_t &= p + \frac{\mathbf{B}^2}{2 \Gamma^2} + \frac{(\mathbf{u} \cdot \mathbf{B})^2}{2}.
\end{aligned}
\end{equation}

We close the set of equations with the Taub-Matthews equation of state, which approximates the Synge EoS of a single species relativistic perfect fluid \citep{Mignone2005}: 
\begin{equation}\label{eq:TM}
h=\frac{5}{2}\mathcal{T}+\sqrt{\frac{9}{4}\mathcal{T}^2+1},
\end{equation}
where $h$ is the specific enthalpy and $\mathcal{T}=p/\rho$ is the temperature. 

As explained above at $t=0$ we have a conical jet, in which the velocity is constant and the density and pressure profiles are, following \cite{KF1997}:
\begin{equation}
\rho_{j}(r,z,t=0) = \rho_{j}(0,z_0,0) \left( \frac{R}{R_0} \right)^{-2} ,   
\end{equation}

\begin{equation}
  P_{j}(r,z,t=0) = P_{j}(0,z_0,0) \left( \frac{R}{R_0} \right)^{-2\gamma},  
\end{equation}
where $r$ is the cylindrical radius (measured from the axis of the cone), $R=\sqrt{r^2+z^2}$ is the spherical radius and  $R_0=z_0$. Furthermore, the transverse transition of all the variables from their values in the jet to the values in the external medium are smoothed to avoid numerical noise at the contact discontinuity, as better specified in Appendix \ref{app:smooth}. The external medium is characterized by density and pressure that
decay with distance from the central object as a power law. The external density profile along $z$ is $\rho_{\text{ext}}(z) =\rho_{ext,0} (z/z_0)^{-\eta}$ and the external pressure profile is $P_{\text{ext}}(z) = P_{ext,0} (z/z_0)^{-\eta}$. To prevent the external pressure profile from dissipating due to the pressure gradient force, a pseudo-gravity body force is applied in the z-direction. This force exactly balances the pressure gradient, ensuring the ambient medium remains in strict hydrostatic equilibrium.
\section{Magnetic field initialization}
\label{app:mag_init}

The magnetic field is initialized using a helical configuration defined in the laboratory frame. Following \cite{sciaccaluga25}, the field components in cylindrical coordinates $(r, \phi, z)$ are given by

\begin{align}
    &B_r = \frac{\alpha B_0}{R^2} \exp\left(-\mathcal{X}^2\right) \frac{r}{R}, \label{eq:Br_cyl}\\
    &B_z = \frac{\alpha B_0}{R^2} \exp\left(-\mathcal{X}^2\right) \frac{z}{R}, \label{eq:Bz_cyl}\\
    &B_\phi =  \Gamma \frac{B_0}{R}  \sqrt{\text{e}^{-2\mathcal{X}^2} -\frac{\psi_\chi}{2 \sin^2\theta}\left[\psi_\chi-\psi_\chi \text{e}^{-2\mathcal{X}^2}+\sqrt{2\pi}\,\rm{erf}\left(\sqrt{2}\mathcal{X}\right)\right]},
\end{align}
where $R = \sqrt{r^2 + z^2}$ is the spherical radius, and $\Gamma$ is the Lorentz factor of the flow. The angular dependence is controlled by the variable $\mathcal{X} = (\cos\theta-1)/\psi_\chi$, where $\theta = \arccos(z/R)$ is the polar angle. The parameter $\psi_\chi = 10^{-2}\theta_j$ sets the decay scale of the field to ensure it vanishes outside the jet.

\begin{figure}[htbp]
\begin{center}
   \includegraphics[width=\linewidth]{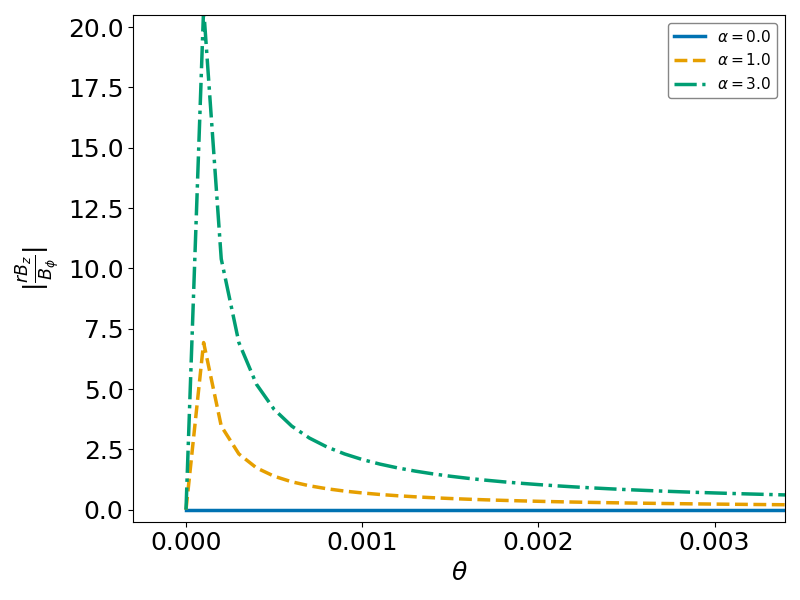}
\caption{Radial profiles of magnetic pitch for three different pitch parameters.}
\label{fig:pitch_sim_frame}
\end{center}
\end{figure}

The parameter $\alpha$ determines the magnetic pitch parameter (the ratio of poloidal to toroidal components). $B_0$ is the reference field amplitude. The magnetization $\sigma$ at the injection point is related to $B_0$ by
\begin{equation}
    \sigma = \frac{B_0^2}{\rho(z_0)}.
\end{equation}
In this setup, $\alpha=1$ implies that the maximum strengths of the poloidal and toroidal components are comparable in the fluid rest frame, whereas in the laboratory frame, the toroidal component dominates by a factor of $\Gamma$. In Fig. \ref{fig:pitch_sim_frame} we present the radial profile of magnetic pitch, for three different values of pitch parameters, as it was defined by \cite{Bodo2021}.

\section{Smoothed profiles and initial conditions}\label{app:smooth}

The simulations adopt smoothed radial profiles to ensure a continuous transition between the jet and ambient medium, thereby minimizing numerical artifacts at the jet boundary, \cite{Boula25}. These profiles are implemented using the hyperbolic secant ($\operatorname{sech}$) function, which allows for smooth yet steep gradients.

The general expression for a smoothed quantity $q$ is given by
\begin{equation}
q(r, z) = q_{\text{ext}} + (q_{\text{j}} - q_{\text{ext}})\, \operatorname{sech}\left[\left(\frac{r}{z \theta_q}\right)^{\alpha_q}\right],
\end{equation}
where $q_{\text{ext}}$ and $q_{\text{j}}$ are the values of the quantity in the ambient medium and jet, respectively; $\theta_q$ sets the angular width of the transition layer, and $\alpha_q$ controls the steepness of the profile.

The cylindrical radial distance $r$ and spherical radius $r'$ are defined as
\begin{align}
r &= \sqrt{x^2 + y^2}, \\
r' &= \sqrt{x^2 + y^2 + z^2}.
\end{align}

The Lorentz factor profile is initialized as
\begin{equation}
\Gamma(r, z) = 1 + (\Gamma_j - 1)\, \operatorname{sech}\left[\left(\frac{r}{z \theta_\Gamma}\right)^{\alpha_\Gamma}\right],
\end{equation}
from which the 3-velocity magnitude is calculated using
\begin{equation}
v_0 = \sqrt{1 - \frac{1}{\Gamma^2}}.
\end{equation}

The density profiles are given by
\begin{align}
\rho_{\text{ext}}(z) &= z^{-\eta}, \\
\rho_j(r') &= \nu\, r'^{-2}, \\
\rho_f(r, z) &= \rho_{\text{ext}} + (\rho_j - \rho_{\text{ext}})\, \operatorname{sech}\left[\left(\frac{r}{z \theta_\rho}\right)^{\alpha_\rho}\right],
\end{align}
where $\nu$ is the density contrast between the jet and the ambient medium, and $\eta$ controls the stratification of the external atmosphere.

Similarly, the pressure profiles follow:
\begin{align}
p_{\text{ext}}(z) &= P_0\, z^{-\eta}, \\
p_j(r') &= P_0\, r'^{-2\gamma}\, P_{\text{ratio}}, \\
p_f(r, z) &= p_{\text{ext}} + (p_j - p_{\text{ext}})\, \operatorname{sech}\left[\left(\frac{r}{z \theta_p}\right)^{\alpha_p}\right],
\end{align}
with $P_{\text{ratio}}$ denoting the jet-to-ambient pressure ratio.

For points inside the jet cone, defined by $r/z < \theta_j$, the velocity components are initialized as
\begin{align}
u_x = v_0\, \frac{x}{r'}, ~
u_y = v_0\, \frac{y}{r'}, ~
u_z = v_0\, \frac{z}{r'}.
\end{align}

The parameters used to define the smoothed profiles and physical conditions in the simulations are: $\theta_j = 0.1$, $\Gamma_j = 10$, $P_0 = 3 \times 10^{-6}$, $\theta_\Gamma = 0.07$, $\alpha_\Gamma = 13$, $\theta_p = 0.13$, $\alpha_p = 15$, $\theta_\rho = 0.1$, $\alpha_\rho = 15$, $\theta_\beta = 0.004$, $b_z = 10^{-5}$, and $\alpha_z = 10^{-10}$.

\section{Density structure of all simulated cases}\label{app:density_map}
\begin{figure*}[htbp]
\begin{center}
   \includegraphics[width=0.9\linewidth]{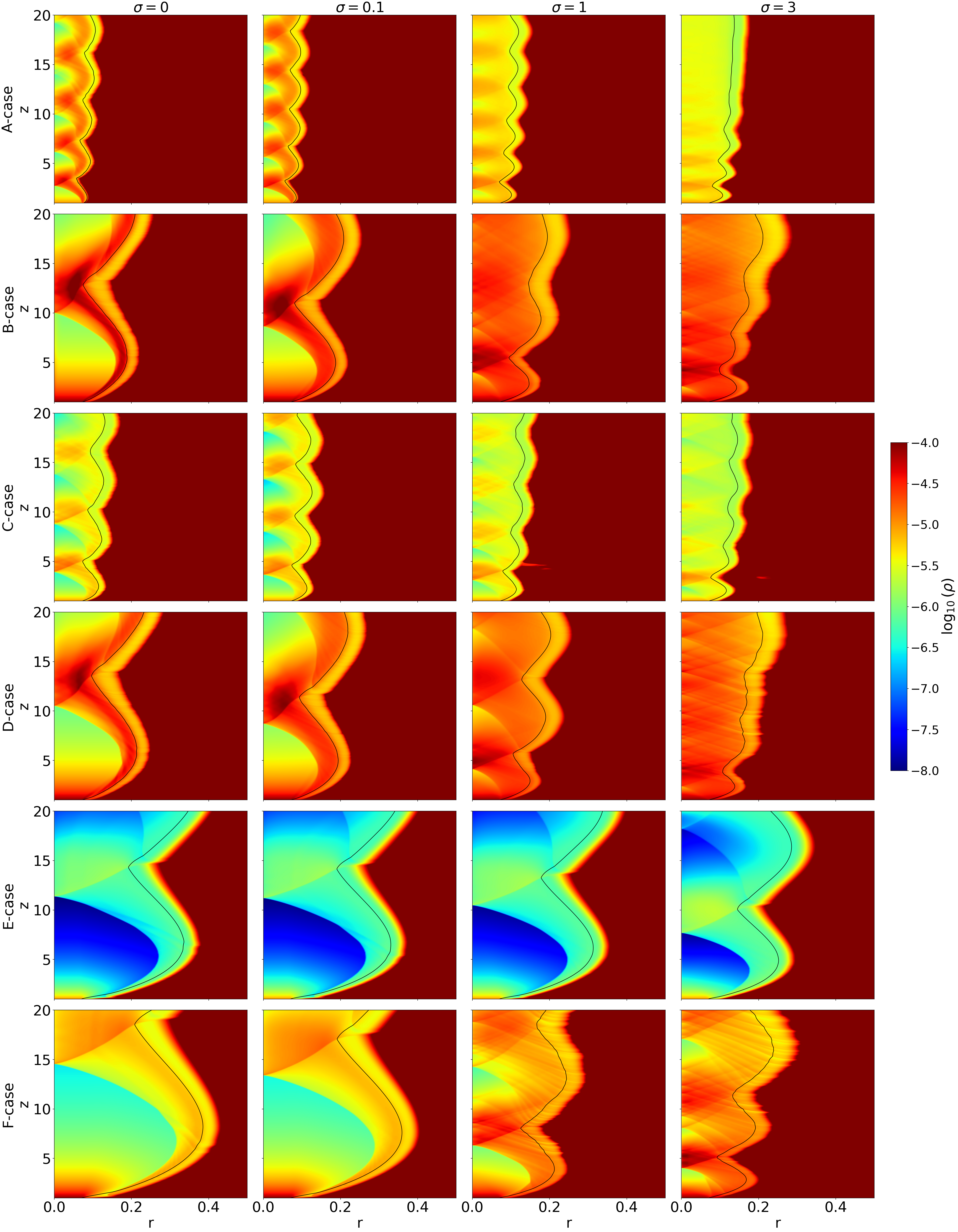}
\caption{
Density maps, $\log_{10}(\rho)$, in the $(x, z)$ plane for all simulated model families (A--F) and for different injection magnetization values ($\sigma = 0, 0.1, 1, 3$). Each row corresponds to a different set of ambient or thermal conditions (Run A to F, as detailed in Table~\ref{tab:parameters}), and each column corresponds to a specific magnetization $\sigma$. The solid black line marks the position of a specific streamline, highlighting the degree of jet expansion and collimation. The color bar represents the decimal logarithm of the density, $\log_{10}(\rho)$. This figure shows the initial differences in the jet's cross-sectional structure, confinement, and the position of the first recollimation point as a function of both the magnetization $\sigma$ and the external medium parameters.\label{fig:density_structure_all}}
\end{center}
\end{figure*}

Figure~\ref{fig:density_structure_all} presents the 2D density maps, $\log_{10}(\rho)$, for all model families (A--F) and all injection magnetizations ($\sigma = 0$ to $\sigma = 3$). This figure clearly illustrates the global structural response of the jets to varying ambient conditions and magnetic field strength. The core features, such as the position of the first recollimation shock and the degree of radial expansion, are visible across the parameter space.

Crucially, the maps for the hydrodynamic cases ($\sigma=0$) and the weakly magnetized cases ($\sigma=0.1$) display highly similar overall geometric structure. Specifically, the maximum radial expansion point ($R_{\max}$) and the axial position of the recollimation shock ($z_{\max}$) show negligible differences between these two columns for a given jet family (A--F). This indicates that magnetic pressure and tension are dynamically insignificant in the $\sigma=0.1$ regime compared to the flow's thermal pressure and inertia. Consequently, the discussion in the main body of the paper focuses on the dynamics in which the magnetic field becomes a dominant physical component, namely in the moderately to highly magnetized regimes ($\sigma \ge 1$ and $\sigma = 3$). These higher-$\sigma$ cases reveal distinct structural differences, such as the systematic shift of $z_{\max}$ towards the injection boundary and the transition to a magnetically-dominated jet core, which are critical for understanding the jet's stability and energy partitioning.

\section{Profiles and parameters}\label{ap:parameters}
\begin{figure}[htbp]
\begin{center}
   \includegraphics[width=\linewidth]{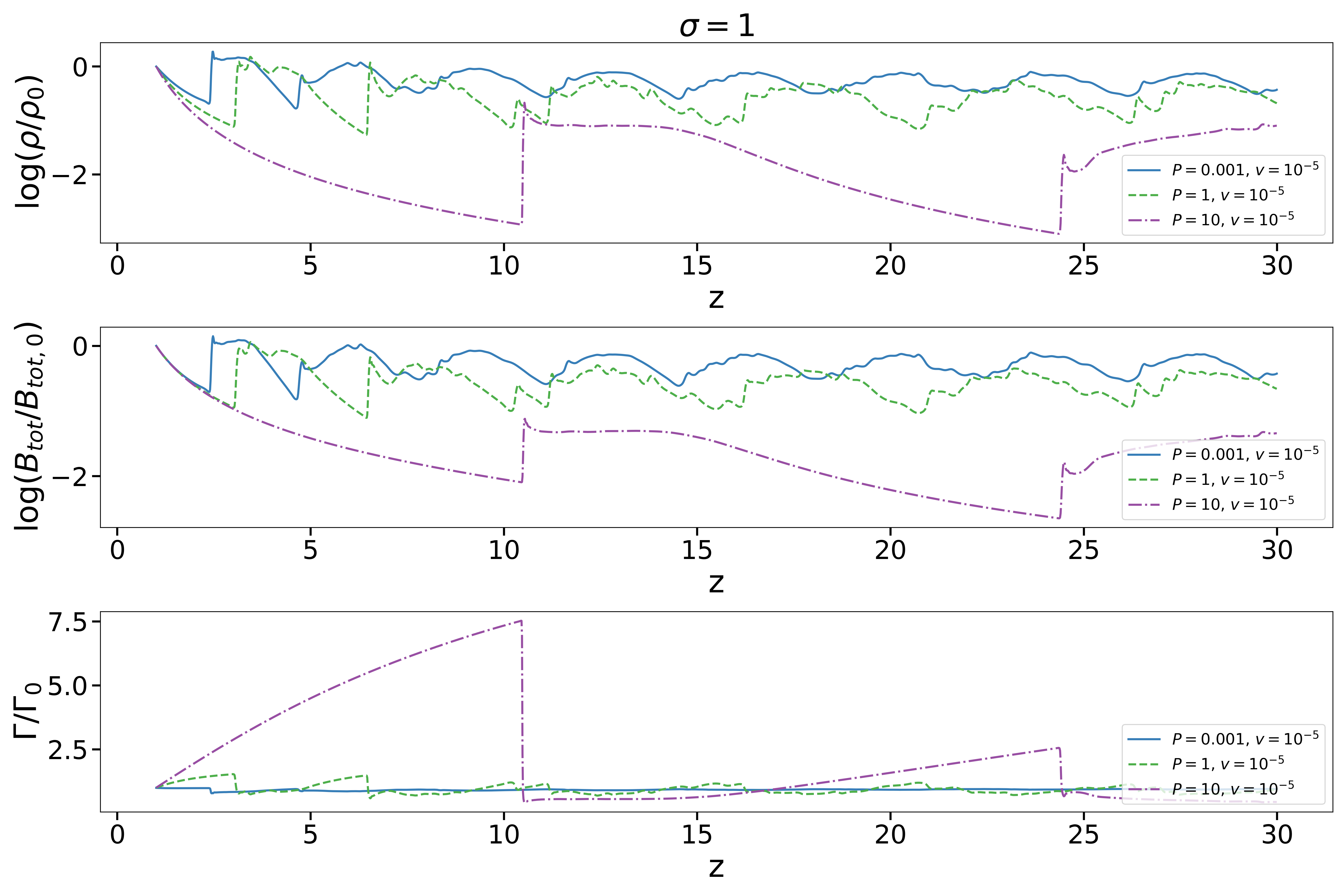}
\caption{Axial profiles of normalized density, total magnetic field, and Lorentz factor for jets with $\sigma=1$. The density ratio $\nu$ is kept constant, while the pressure ratio varies. The three subplots illustrate how changes in pressure affect the jet structure, magnetic field strength, and acceleration along the axis.}
\label{fig:jet_sigma1_constant_nu_profiles}
\end{center}
\end{figure}

\begin{figure}[htbp]
\begin{center}
   \includegraphics[width=\linewidth]{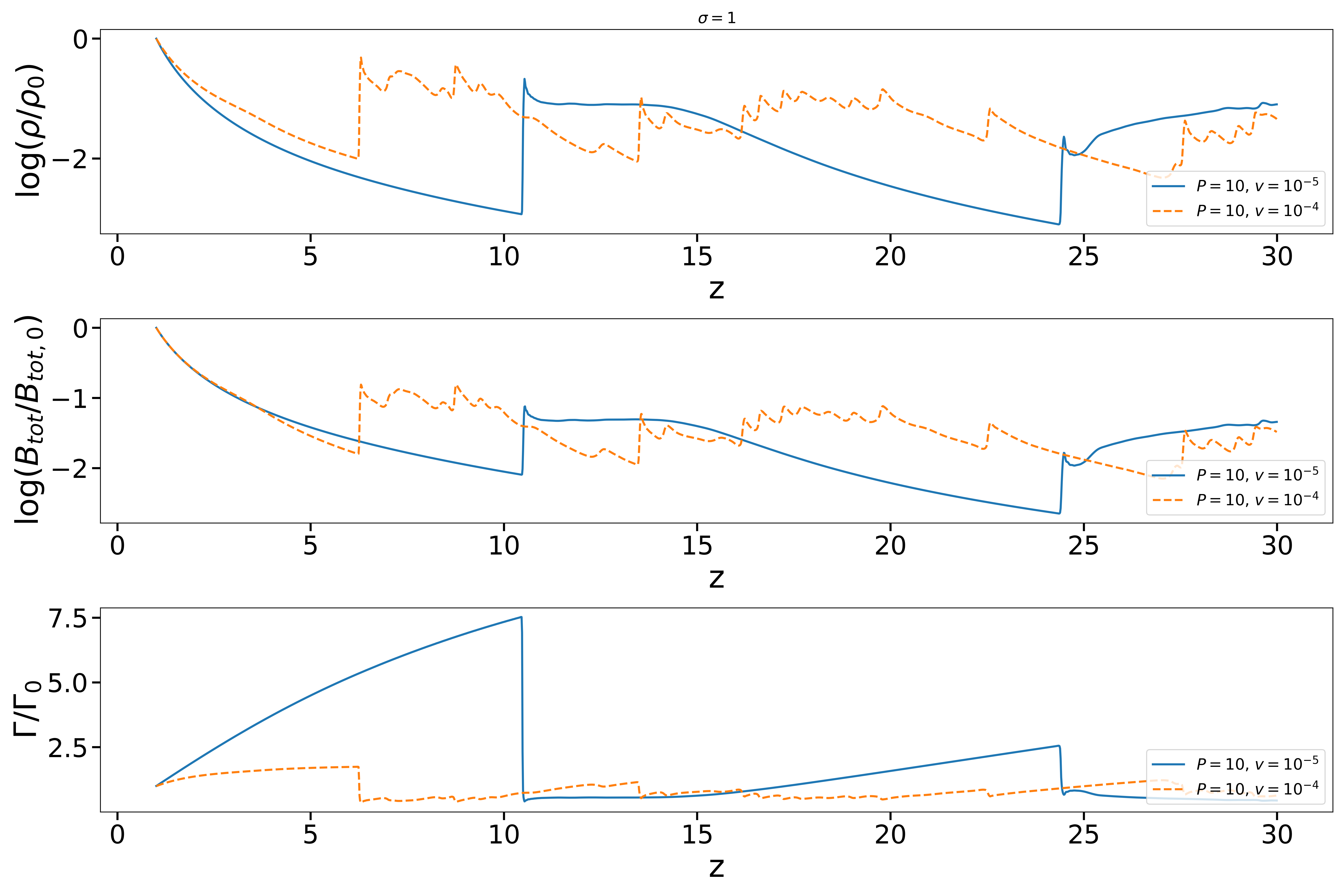}
\caption{Axial profiles of normalized density, total magnetic field, and Lorentz factor for jets with $\sigma=1$ in the hot jet case (pressure ratio 10). Two different density ratios $\nu$ are considered, showing the impact of density contrast on the jet structure, magnetic field, and acceleration along the axis.}
\label{fig:jet_sigma1_hot_vary_nu_profiles}
\end{center}
\end{figure}

Figures~\ref{fig:jet_sigma1_constant_nu_profiles} and~\ref{fig:jet_sigma1_hot_vary_nu_profiles} further explore the impact of the initial pressure ratio on the steady-state structure for a fixed magnetization of $\sigma=1$. Figure~\ref{fig:jet_sigma1_constant_nu_profiles} compares models with constant density ratio ($\nu=10^{-5}$) but varying pressure $P$. It demonstrates that only the highly over-pressured jet ($P=10$; dash-dotted purple line) achieves significant acceleration, converting magnetic energy into kinetic energy to reach a peak Lorentz factor of $\Gamma/\Gamma_0 \approx 7.5$. In contrast, the lower-pressure cases ($P=0.001$ and $P=1$) exhibit minimal acceleration ($\Gamma/\Gamma_0 \approx 1$). The density and magnetic field profiles for the $P=10$ case show a steeper initial decay due to greater expansion, followed by a strong recollimation shock.

In Fig.~\ref{fig:jet_sigma1_hot_vary_nu_profiles}, the pressure ratio is constant at $P=10$ (a hot jet) while we vary the density ratio $\nu$ and increase $\nu$ from $10^{-5}$ to $10^{-4}$ (dashed orange line). A substantial shift in the recollimation indicates increased confinement, with shocks closer to the injection boundary (e.g., the first shock moves from $z \approx 10$ to $z \approx 7$). This earlier recollimation also limits the maximum flow acceleration, reducing the peak Lorentz factor from $\Gamma/\Gamma_0 \approx 7.5$ ($\nu=10^{-5}$) to $\Gamma/\Gamma_0 \approx 2.0$ ($\nu=10^{-4}$), as the energy conversion is cut short by the shock. The continuous acceleration observed in light jets is driven by the conversion of high specific enthalpy (thermal energy) into bulk kinetic energy, while the confinement and the resulting recollimation shock location are determined by the equilibration of the jet's internal pressure against the ambient medium.

\subsection{Normalization and energy budget}\label{Ub}
To perform a consistent parameter study, we must ensure that varying the pitch parameter $\alpha$ changes the field topology without altering the total power budget of the jet. We achieve this by normalizing the field amplitude $B_0(\alpha)$ such that the cross-sectional integrated magnetic energy matches the target magnetization $\sigma$.

The comoving magnetic energy density $U'_B$ is defined as
\begin{equation}
    U'_B(R, \theta) = \frac{B'_\phi{}^2}{8\pi} + \frac{B'_{pol}{}^2}{8\pi} \propto \frac{B_\phi^2}{\Gamma^2} + \left(B_r^2 + B_z^2\right).
\end{equation}
Substituting the field expressions from Eqs. \ref{eq:Br_cyl} and \ref{eq:Bz_cyl}, this becomes
\begin{equation}
    U'_B(R, \theta) \propto \frac{B_0^2}{R^2} \left[ f_t(\theta) + \frac{\alpha^2}{R^2} f_p^2(\theta) \right],
\end{equation}
where $f_t(\theta)$ is the term under the square root in the $B_\phi$ definition, and $f_p(\theta) = \exp(-\mathcal{X}^2)$.

To determine the normalization constant $B_0(\alpha)$, we integrate this energy density over the injection surface (a spherical cap at $R=R_0$) and equate it to the target total magnetic energy $E_{mag} \propto \sigma$:
\begin{equation}
    B_0(\alpha) {\displaystyle \int_0^{\theta_{\max}} \left( f_t(\theta) + \frac{\alpha^2}{R_0^2} f_p^2(\theta) \right) \sin\theta \, d\theta} = constant
\end{equation}
This normalization ensures that as we vary $\alpha$, the amplitude $B_0$ adjusts automatically to maintain a constant total magnetic energy flux injection.

\section{Pitch parameter and equilibrium}\label{app:pitch}

\begin{figure}[htbp]
\begin{center}
\includegraphics[width=\linewidth]{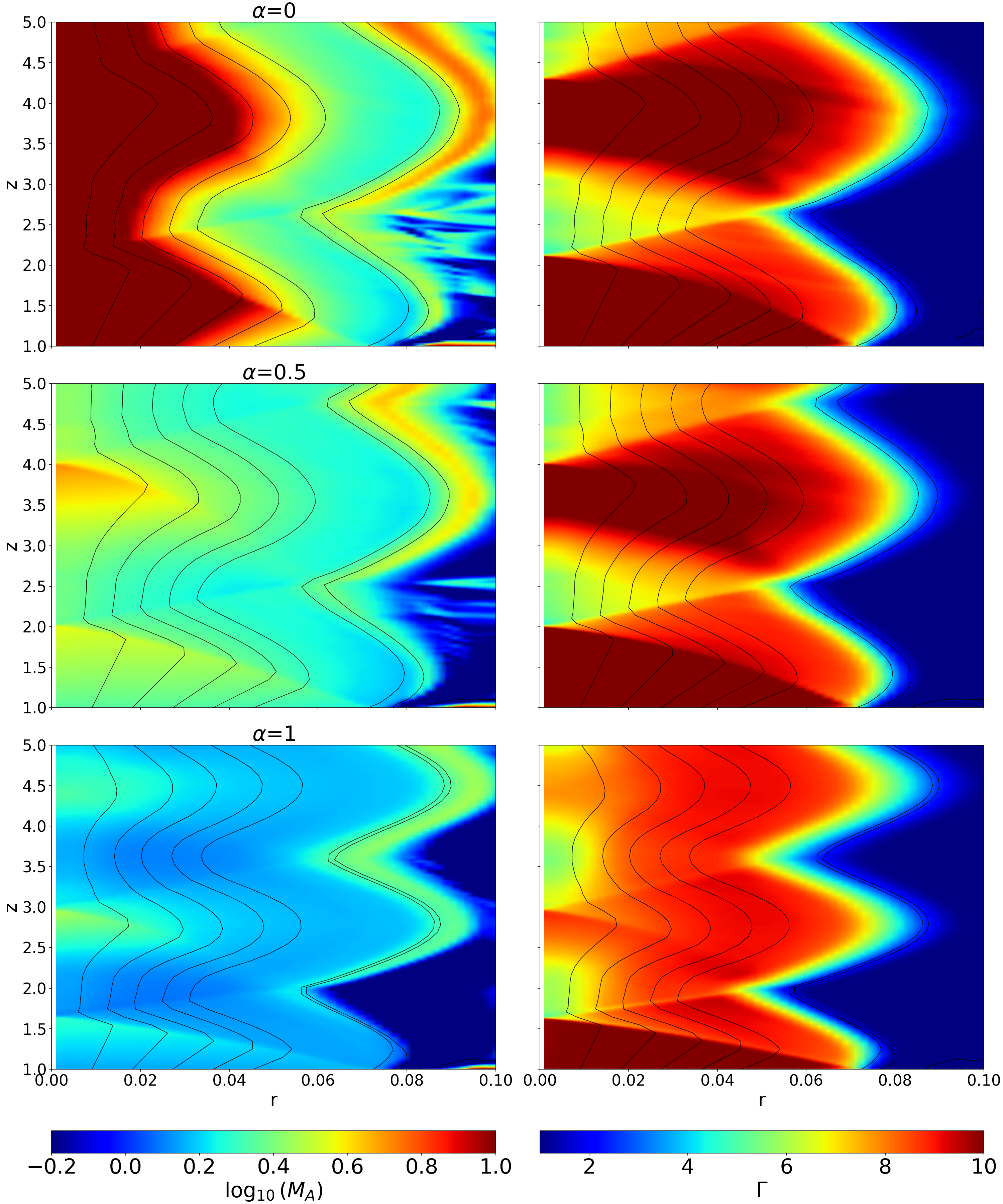}
\caption{2D maps of the base-10 logarithm of the Alfvénic Mach number ($\log_{10} M_A$; \textit{left column}) and the Lorentz factor ($\Gamma$; \textit{right} \textit{column}) in the $(x, z)$ plane for the $\sigma=1$ jet, comparing three magnetic pitches: $\alpha=0$ (purely toroidal: \textit{top}), $\alpha=0.5$ (\textit{middle}), and $\alpha=1$ (equal components; \textit{bottom}). \label{fig:pitch_maps}}
\end{center}
\end{figure}

\begin{figure}[htbp]
\begin{center}
\includegraphics[width=\linewidth]{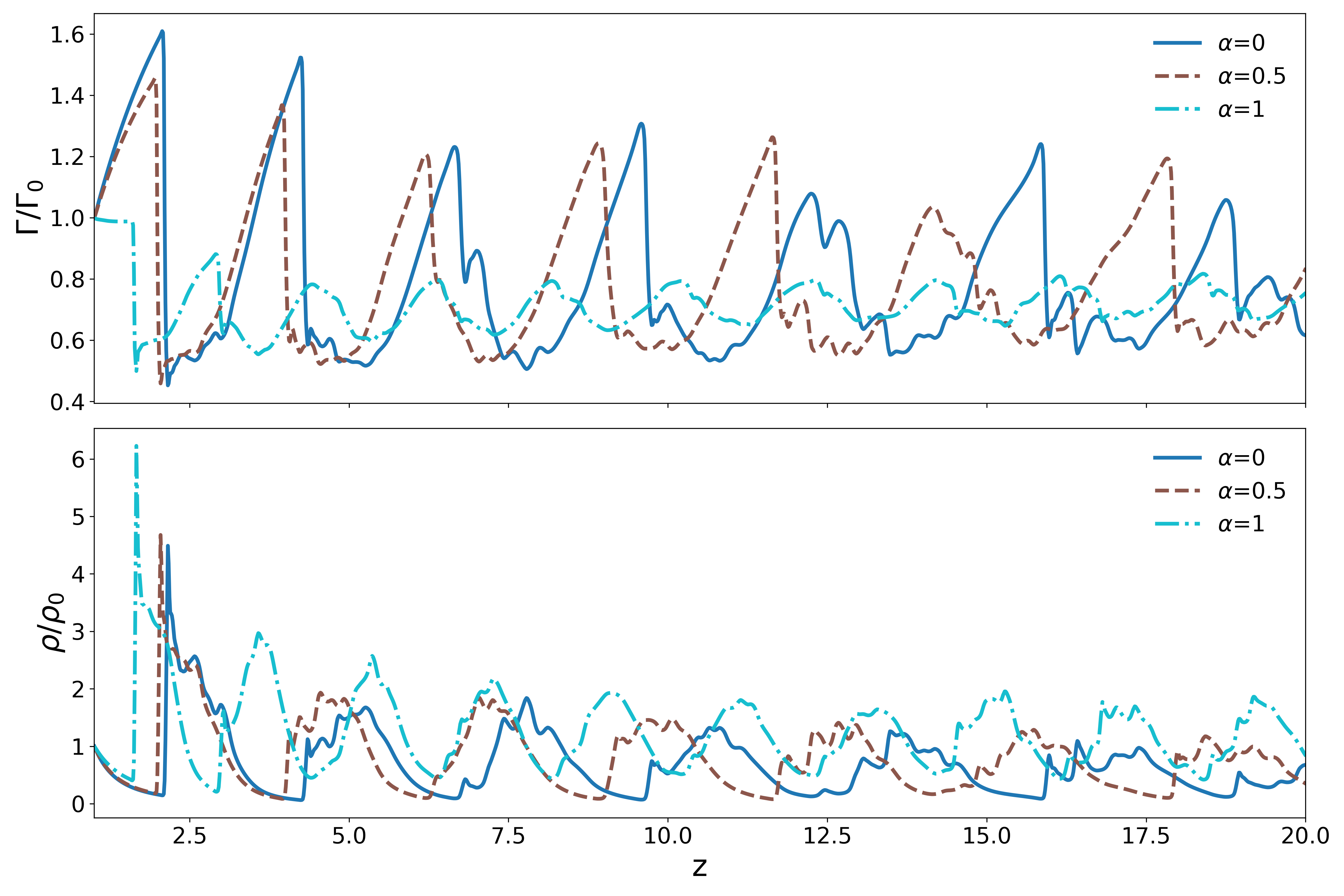}
\caption{Axial profiles of the normalized Lorentz factor ($\Gamma/\Gamma_0$; \textit{top}) and normalized density ($\rho/\rho_0$; \textit{bottom}) along the jet axis ($x=0$), comparing three magnetic field pitch: $\alpha=0$ (solid blue), $\alpha=0.5$ (dashed brown), and $\alpha=1$ (dash-dotted cyan). \label{fig:pitch_prof}}
\end{center}
\end{figure}

\begin{figure}[htbp]
\begin{center}
\includegraphics[width=0.65\linewidth]{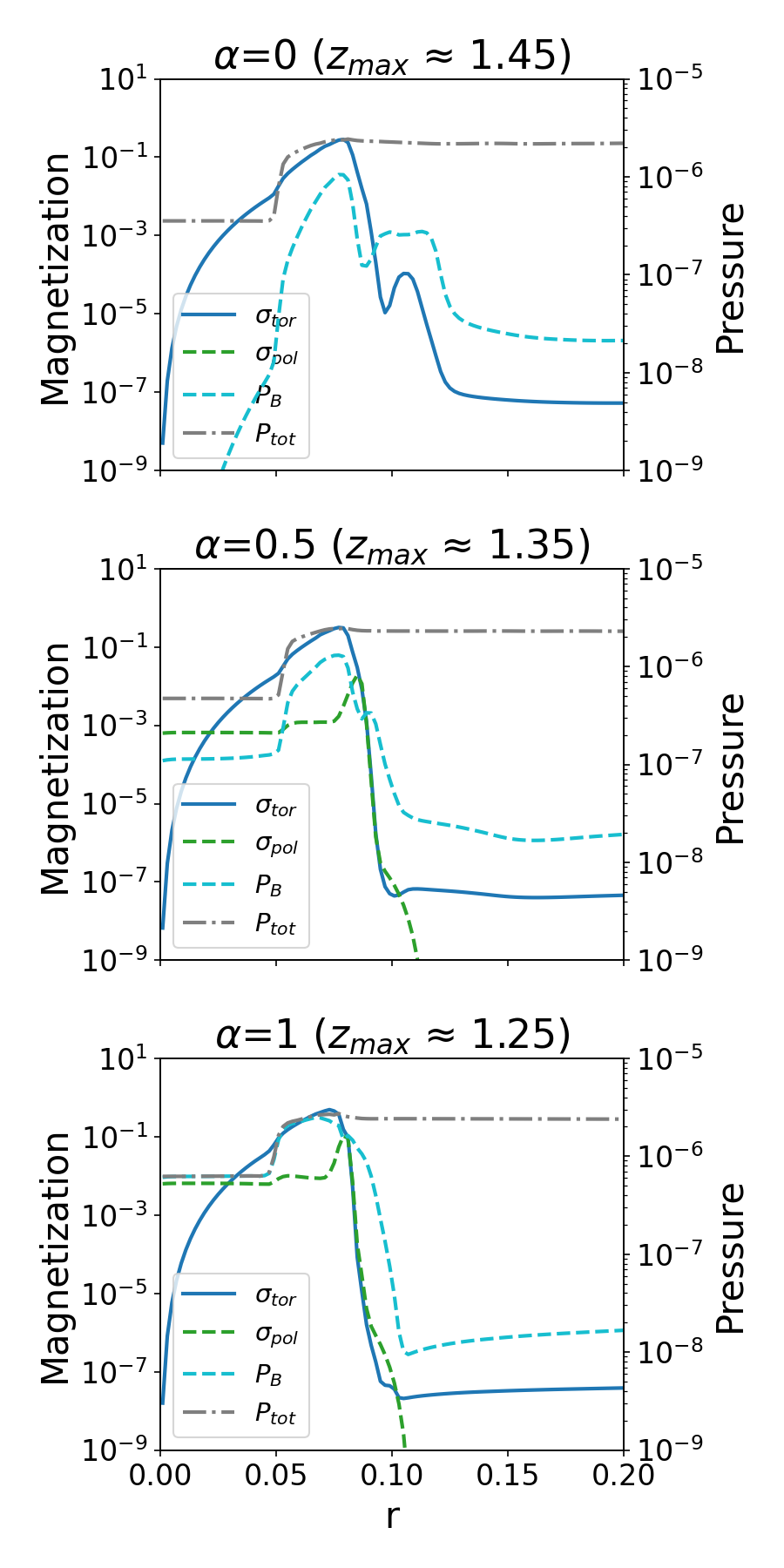}
\caption{Radial profiles of the toroidal magnetization ($\sigma_{\text{tor}}$), poloidal magnetization ($\sigma_{\text{pol}}$), magnetic pressure ($P_B$), and total pressure ($P_{\text{tot}}$) for the three pitch parameter cases, extracted at the position of maximum radial expansion, $z_{\max}$. The $z_{\max}$ value for each case is noted above the corresponding panel. \label{fig:pitch_radial}}
\end{center}
\end{figure}
\begin{figure*}[htbp]
\begin{center}
\includegraphics[width=\linewidth]{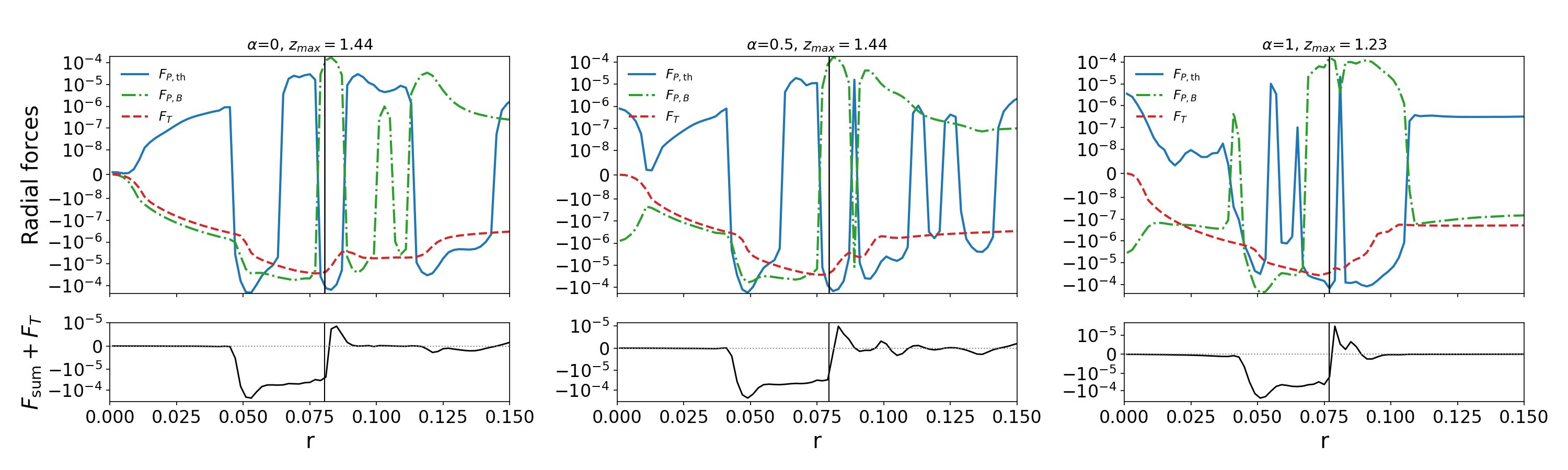}
  \caption{
   Radial force balance analysis for increasing magnetic pitch, from purely toroidal ($\alpha=0$, left) to poloidal-dominated ($\alpha=1$, right). The location of maximum radial expansion ($r_{max}, z_{max}$).
   {Sub-panels:} \textit{Top panels}: Pressure gradient ($F_{P,th} + F_{P,B}$, blue) vs. magnetic tension ($F_T$; dashed  red). \textit{Bottom panels}: Net radial force. \label{fig:pitch_radial}}
\end{center}
\end{figure*}

Figure~\ref{fig:pitch_maps} illustrates the 2D structure of the jet for different pitches in the case of equilibrium, where the pressure is $p_j = p_{f} + (1 - \alpha^2) {B_0^2}/{2} e^{-2\mathcal{X}^2} r^{2\gamma_{\mathrm{ad}}}$. The $B_{pol}/B_{\phi}=0$ case (purely toroidal field) shows a well-defined, oscillating jet structure, similar to the standard $\sigma=1$ run, with the first recollimation shock clearly visible around $z \approx 1.5$. As the poloidal component is increased ($\alpha$ $0.5$ and $1$), the flow becomes slightly less confined in the region near the injection point, and the jet boundary appears smoother. This suggests that introducing a stronger poloidal field reduces the efficiency of the magnetic hoop-stress mechanism near the engine. $M_A$ maps show that the jet core remains highly sub-Alfvénic ($M_A \ll 1$) in all three cases, consistent with the high initial magnetization of $\sigma=1$. The Lorentz factor maps show a strong acceleration region before the recollimation shock in all topologies, indicating that the magnetic field is efficiently converting its energy into bulk motion, regardless of the precise partitioning between $B_{pol}$ and $B_{\phi}$.

Figure~\ref{fig:pitch_prof} shows the corresponding axial profiles of the normalized Lorentz factor and density. The primary impact of increasing the pitch is on the periodicity and damping of the recollimation shocks. In the purely toroidal case ($B_{pol}/B_{\phi}=0$), the jet exhibits strong, regular oscillations in both $\Gamma$ and $\rho$. Introducing a poloidal field ($B_{pol}/B_{\phi}=0.5$ and $B_{pol}/B_{\phi}=1$) causes the oscillations to damp out more quickly and results in a slight shift of the first recollimation shock to larger distances. However, the difference is minimal ($z \approx 1.7$ for $\alpha$ 0 vs. $z \approx 1.8$ for $\alpha$ 1). The peak density compression at the first shock is highest in the $B_{pol}/B_{\phi}=1$ case, reaching $\rho/\rho_0 \approx 5$, compared to $\rho/\rho_0 \approx 4.5$ in the $B_{pol}/B_{\phi}=0$ case. The final structure appears more stable and less oscillatory for the higher pitch.

Figure~\ref{fig:pitch_radial} provides insight into the local pressure balance by showing the radial profiles of magnetization and pressure extracted near the first recollimation point. For $B_{pol}/B_{\phi}=0$, $\sigma_{\text{tor}}$ is entirely responsible for the magnetic pressure, dominating $\sigma_{\text{pol}}$ by orders of magnitude. As the pitch parameter increases, $\sigma_{\text{pol}}$ grows significantly, approaching the magnitude of $\sigma_{\text{tor}}$ in the $B_{pol}/B_{\phi}=1$ case. However, the total magnetic pressure and total pressure exhibit similar magnitudes and profiles across the three cases, indicating that the total pressure available for confinement is preserved. The main difference lies in the width of the magnetic confinement layer: the $B_{pol}/B_{\phi}=0$ case shows a broader pressure peak, while the $B_{pol}/B_{\phi}=1$ case results in a slightly narrower jet, consistent with the observed $z_{\max}$ values listed in the panels, where higher pitch parameter corresponds to a smaller $z_{\max}$. This suggests that while a purely toroidal field provides strong hoop stress, a mixed field topology can achieve comparable total pressure confinement, leading to similar global stability, but may alter the damping rate of axial oscillations.

\section{Linear analysis of the centrifugal instability}\label{sec:linear}
\begin{figure}[htbp]
 \begin{center} \includegraphics[width=0.65\linewidth]{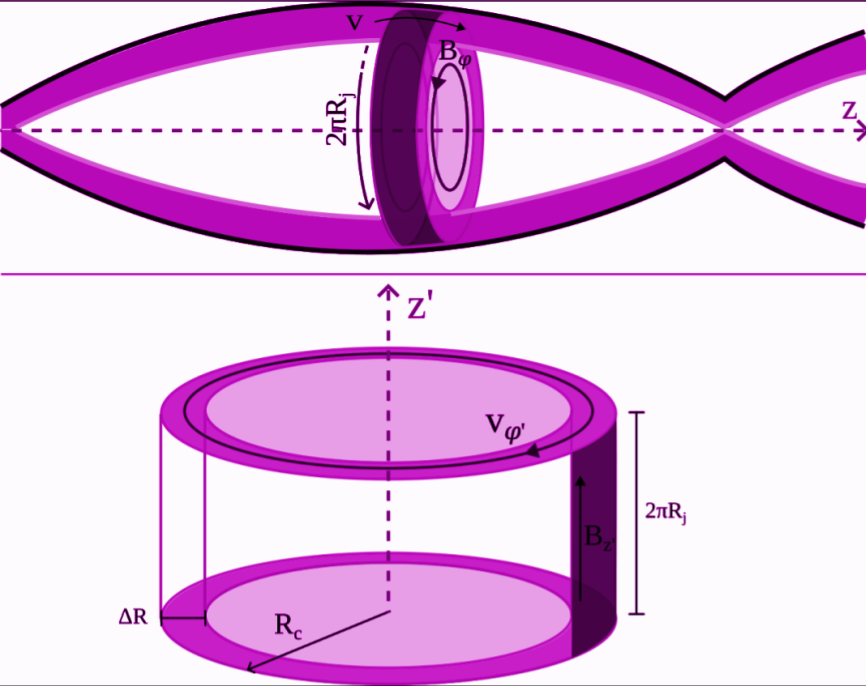} \caption{Cartoon illustrating the configurations used in the linear analysis, adapted from \cite{Komissarov2019} and \cite{Boula25}. {\textit{Top}:} Relativistic jet with a purely azimuthal magnetic field undergoes recollimation due to the confining action of an external pressure. The white inner region represents the freely expanding flow, while the shaded area corresponds to the shocked outer layer. {\textit{Bottom}:} Magnetised, rotating cylindrical shell with a purely axial magnetic field, representing the idealised configuration analysed in this work.\label{fig:CFI_c}} \end{center}
 \end{figure}

 We examine the stability of a simplified, localized configuration to compute the growth rate of the CFI, similar to the approach used by \cite{Komissarov2019} and was presented in \cite{Boula25}. This is a linear stability analysis, following the methodology detailed in \cite{Vlahakis2023}. We model the jet as a magnetized, rotating cylindrical shell with a purely axial magnetic field, distinct from the recollimating jet structure analyzed in the main text (Fig. \ref{fig:CFI_c} illustrates this idealization, adapted from \citealt{Komissarov2019} and \citealt{Boula25}). We adopt a cylindrical coordinate system $(R, z)$ centered on the shell axis. The shell has outer radius $R_{\text{c}}$, width $\Delta R$, constant azimuthal velocity $V_\phi$ (Lorentz factor $\Gamma$), uniform specific enthalpy $h$, and magnetization $\sigma$. The magnetic field is $\mathbf{b}=\sqrt{\sigma h\rho}\hat{z}$, and the thermal pressure is $P = \rho \mathcal{T}$. Radial force balance inside the shell mandates a power-law density profile $\rho(R) = \rho_j (R/R_{\text{c}})^a$, with the exponent defined by $a = (\Gamma^2 - 1)(1+\sigma)/({\mathcal{T}}/{h} + {\sigma}/{2})$ We fix the shell width as $\Delta R = R_{\text{c}}/a$. The shell is embedded in an external medium characterized by constant properties (${\rho_{\text{ext}}}$, $h_{\text{ext}}$, and axial magnetic field $\sqrt{\sigma_{\text{ext}} h_{\text{ext}} \rho_{\text{ext}}}$). Pressure equilibrium at the shell boundary defines the density ratio: $$\frac{\rho_{\text{ext}}}{\rho_{R_c}} = \frac{\mathcal{T}_{R_c} + \sigma_{R_c} h_{R_c}/2}{\mathcal{T}_{\text{ext}} + \sigma_{\text{ext}} h_{\text{ext}}/2}$$ The external-to-internal specific enthalpy ratio, $h_e \equiv h_{\text{ext}}/h$, controls the density contrast, with significant effects appearing only when $h_e \gg 1$. Given that the CFI eigenfunction is localized near the shell’s outer edge, we treat the inner boundary as a rigid wall. We introduce perturbations of the form $f(R) e^{i(kz - \omega t)}$ into the linearized equations. Solving across the shell and the external region using appropriate matching conditions \cite{Vlahakis2023} yields the eigenvalue problem. The solution provides the (purely imaginary) instability growth rate $\Im \omega$ as a function of the wavenumber $k$.  The results are consistent with the qualitative criterion proposed by \cite{Komissarov2019}, where the growth of the CFI is linked to the ratio $(\sigma_{\mathrm{tor}}/(\sigma+1))/\Gamma^2 \sim \sigma_{\mathrm{tor}}/\Gamma^2$. 

\begin{table}[h!]
\centering
\scriptsize
\caption{Summary of $\sigma_{tor}/\Gamma^2$, curvature, and related quantities for all cases.}
\begin{tabular}{@{}c c c c c c c@{}}
\hline
Case & Stream. & $\sigma_{tor}/\Gamma^2$ & $x_{\rm max}$ & $R_{\rm curv}$ & $x/R$ & $\sigma_{tor}/(x/R)$ \\
\hline
A2 & 0.075 & 1.805e-03 & 0.0798 & 10.945 & 7.30e-03 & 2.475e-01 \\
A2 & 0.085 & 5.392e-07 & 0.0870 & 0.2062 & 4.219e-01 & 1.278e-06 \\
A3 & 0.075 & 8.435e-03 & 0.0948 & 9.333 & 1.02e-02 & 8.301e-01 \\
A3 & 0.085 & 1.485e-02 & 0.1064 & 12.422 & 8.60e-03 & 1.733 \\
A4 & 0.075 & 1.439e-02 & 0.1003 & 6.500 & 1.54e-02 & 9.323e-01 \\
A4 & 0.085 & 2.898e-02 & 0.1124 & 7.202 & 1.56e-02 & 1.857 \\
B2 & 0.075 & 9.110e-04 & 0.1800 & 59.906 & 3.00e-03 & 3.032e-01 \\
B2 & 0.085 & 9.137e-04 & 0.1929 & 84.185 & 2.30e-03 & 3.987e-01 \\
B3 & 0.075 & 5.822e-03 & 0.1391 & 26.065 & 5.30e-03 & 1.091 \\
B3 & 0.085 & 9.850e-03 & 0.1542 & 48.572 & 3.20e-03 & 3.103 \\
B4 & 0.075 & 9.981e-03 & 0.1355 & 18.755 & 7.20e-03 & 1.381 \\
B4 & 0.085 & 1.918e-02 & 0.1512 & 15.417 & 9.80e-03 & 1.955 \\
C2 & 0.075 & 4.193e-04 & 0.1087 & 22.164 & 4.90e-03 & 8.553e-02 \\
C2 & 0.085 & 6.085e-04 & 0.1173 & 18.346 & 6.40e-03 & 9.516e-02 \\
C3 & 0.075 & 3.096e-03 & 0.1080 & 10.909 & 9.90e-03 & 0.313 \\
C3 & 0.085 & 5.924e-03 & 0.1189 & 10.556 & 1.13e-02 & 0.526 \\
C4 & 0.075 & 7.859e-03 & 0.1101 & 13.021 & 8.50e-03 & 0.930 \\
C4 & 0.085 & 1.606e-02 & 0.1220 & 13.635 & 8.90e-03 & 1.795 \\
D2 & 0.075 & 8.029e-04 & 0.1874 & 86.514 & 2.20e-03 & 0.371 \\
D2 & 0.085 & 8.476e-04 & 0.1998 & 50.267 & 4.00e-03 & 0.213 \\
D3 & 0.075 & 5.038e-03 & 0.1417 & 24.884 & 5.70e-03 & 0.884 \\
D3 & 0.085 & 8.716e-03 & 0.1565 & 28.297 & 5.50e-03 & 1.575 \\
D4 & 0.075 & 1.238e-02 & 0.1303 & 16.126 & 8.10e-03 & 1.531 \\
D4 & 0.085 & 2.223e-02 & 0.1456 & 14.167 & 1.03e-02 & 2.163 \\
E2 & 0.075 & 3.846e-05 & 0.3262 & 114.875 & 2.80e-03 & 1.354e-02 \\
E2 & 0.085 & 6.611e-05 & 0.3334 & 120.801 & 2.80e-03 & 2.395e-02 \\
E3 & 0.075 & 3.639e-04 & 0.3090 & 90.745 & 3.40e-03 & 0.107 \\
E3 & 0.085 & 6.353e-04 & 0.3167 & 102.875 & 3.10e-03 & 0.206 \\
E4 & 0.075 & 9.271e-04 & 0.2803 & 66.154 & 4.20e-03 & 0.219 \\
E4 & 0.085 & 1.644e-03 & 0.2885 & 72.655 & 4.00e-03 & 0.414 \\
F2 & 0.075 & 4.069e-04 & 0.3539 & 103.643 & 3.40e-03 & 0.119 \\
F2 & 0.085 & 6.263e-04 & 0.3635 & 104.452 & 3.50e-03 & 0.180 \\
F3 & 0.075 & 2.993e-03 & 0.2075 & 47.579 & 4.40e-03 & 0.686 \\
F3 & 0.085 & 5.139e-03 & 0.2168 & 43.865 & 4.90e-03 & 1.040 \\
F4 & 0.075 & 7.393e-03 & 0.1618 & 19.370 & 8.40e-03 & 0.885 \\
F4 & 0.085 & 1.469e-02 & 0.1756 & 18.227 & 9.60e-03 & 1.525 \\
\hline
\end{tabular}
\label{tab:sigma_curvature_compact}
\end{table}

\end{appendix}

\end{document}